\documentclass[a4paper,fleqn,usenatbib,useAMS]{mnras}

\usepackage[T1]{fontenc}
\usepackage{ae,aecompl}

\usepackage{graphicx}	
\usepackage{amsmath}	
\usepackage{amssymb}	
\usepackage{float}
\usepackage{mathptmx}
\usepackage{hyperref}

\newcommand\lsun{\rm L_{\odot}}
\newcommand\msun{\rm M_{\odot}}

\newcommand\hi{H\,{\sc i}$\;$}
\newcommand\htwo{\rm H_2}

\newcommand\co{$^{12}{\rm CO}$}
\newcommand\colong{$^{12}{\rm CO}\;{\rm J}\;=1\rightarrow0\;$}
\newcommand\coshort{${\rm CO}\;$}

\title[COOL BUDHIES I]{Early Science with the Large Millimeter Telescope: COOL BUDHIES I - a pilot study of molecular and atomic gas at $z\simeq0.2$}

\author[R.~Cybulski et al.]{Ryan Cybulski,$^{1}$\thanks{E-mail:
jcybulsk@astro.umass.edu} Min ~S.~Yun,$^{1}$, Neal Erickson$^{1}$, Victor De la Luz$^{2}$, Gopal Narayanan$^{1}$, \newauthor Alfredo Monta\~{n}a$^{3,4}$, David S\'{a}nchez-Arg\"{u}lles$^{3}$, Jorge A.~Zavala$^{3}$, Milagros Zeballos$^{3}$,\newauthor Aeree Chung$^{5}$, Ximena Fern\'{a}ndez$^{6,7}$, Jacqueline van Gorkom$^{6}$, Chris P.~Haines$^{8}$,\newauthor Yara L.~Jaff\'{e}$^{9}$,  Mar\'{i}a Montero-Casta\~{n}o$^{10}$, Bianca M.~Poggianti$^{11}$,\newauthor Marc A.~W.~Verheijen$^{12}$, Hyein Yoon$^{5}$, Kevin Harrington$^{1}$, David H.~Hughes$^{3}$,\newauthor Glenn E.~Morrison$^{13,14}$, F.~Peter Schloerb$^{1}$, Miguel Velazquez$^{3}$\\
$^{1}$Department of Astronomy, University of Massachusetts, Amherst, MA 01003, USA\\
$^{2}$SCiESMEX, Instituto de Geof\'{i}sica, Unidad Michoacan, Universidad Nacional Aut\'{o}noma de M\'{e}xico, Morelia, Michoacan, M\'{e}xico, CP 58190\\
$^{3}$Instituto Nacional de Astrof\'{i}sica, \'{O}ptica, y Electr\'{o}nica (INAOE), Tonantzintla, Luis Enrique Erro 1, Sta. Ma. Tonantzintla, Puebla, M\'{e}xico\\
$^{4}$Consejo Nacional de Ciencia y Tecnolog\'{i}a, Av. Insurgentes Sur 1582, Col. Credito Constructor, Del. Benito Ju\'{a}rez, C.P.: 03940, M\'{e}xico, D.F.\\
$^{5}$Department of Astronomy, Yonsei University, 50 Yonsei-ro Seodaemun-gu, Seoul 120-749, Korea\\
$^{6}$Department of Astronomy, Columbia University, Mail Code 5246, 550 W 120th St., New York, NY 10027, USA\\
$^{7}$Department of Physics and Astronomy, Rutgers University, 136 Frelinghuysen Road, Piscataway, NJ 08854, USA\\
$^{8}$Departamento de Astronom\'{i}a, Universidad de Chile, Casilla 36-D, Correo Central, Santiago, Chile\\
$^{9}$Department of Astronomy, Universidad de Concepci\'{o}n, Casilla 160-C, Concepci\'{o}n, Chile\\
$^{10}$Dunlap Institute for Astronomy and Astrophysics, University of Toronto, 50 St George Street, Toronto, ON M5S 3H4, Canada\\
$^{11}$INAF - Osservatorio Astronomico di Padova, vicolo dell Osservatorio 5, I-35122, Padova, Italy\\
$^{12}$University of Groningen, Kapteyn Astronomical Institute, Landleven 12, 9747 AD, Groningen, The Netherlands\\
$^{13}$Institute for Astronomy, University of Hawaii, Manoa, Hawaii 96822-1897 USA\\
$^{14}$Canada-France-Hawaii Telescope Corp., Kamuela, Hawaii 96743-8432, USA}
\date{Accepted 2015 xxxx. Received 2015 xxxx; in original form 2015 xxxx}

\pubyear{2015}

\begin{document}
\label{firstpage}
\pagerange{\pageref{firstpage}--\pageref{lastpage}}
\maketitle

\begin{abstract}
An understanding of the mass build-up in galaxies over time necessitates tracing the evolution of cold gas (molecular and atomic) in galaxies. To that end, we have conducted a pilot study called CO Observations with the LMT of the Blind Ultra-Deep \hi Environment Survey (COOL BUDHIES). We have observed 23 galaxies in and around the two clusters Abell 2192 ($z=0.188$) and Abell 963 ($z=0.206$), where 12 are cluster members and 11 are slightly in the foreground or background, using about 28 total hours on the Redshift Search Receiver (RSR) on the Large Millimeter Telescope (LMT) to measure the \colong emission line and obtain molecular gas masses. These new observations provide a unique opportunity to probe {\it both} the molecular and atomic components of galaxies as a function of environment beyond the local Universe. For our sample of 23 galaxies, nine have reliable detections (${\rm S/N}\geq3.6$) of the \co$\;$ line, and another six have marginal detections ($2.0<{\rm S/N}<3.6$). For the remaining eight targets we can place upper limits on molecular gas masses roughly between $10^9$ and $10^{10} \msun$. Comparing our results to other studies of molecular gas, we find that our sample is significantly more abundant in molecular gas overall, when compared to the stellar and the atomic gas component, and our median molecular gas fraction lies about $1\sigma$ above the upper limits of proposed redshift evolution in earlier studies. We discuss possible reasons for this discrepancy, with the most likely conclusion being target selection and Eddington bias.
\end{abstract}




\section{Introduction}
One of the most important goals of modern astrophysics is to understand how galaxies have evolved over cosmic time. One can approach this goal by examining the morphologies, stellar mass build-up, colours, and star-formation histories of galaxies as a function of redshift. A number of studies along these lines have also revealed that the properties of galaxies strongly depend on local environment, as galaxies residing in regions of higher density at $z\leq1$ are more frequently massive, early-type, and passively-evolving \citep[e.g.,][]{dressler1980, treu2003, poggianti2006, haines2007, gallazzi2009, tran2009, gavazzi2010, mahajan2010, jaffe2011, rasmussen2012, scoville2013, cybulski2014}. However, it is also fundamentally important to examine how the evolution of gas in the interstellar medium (ISM) of galaxies impacts the growth of stellar mass over cosmic time and as a function of environment. A key observational tool for these efforts is the cold gas content of galaxies, both the atomic (\hi) and the molecular ($\htwo$, commonly traced by the line emission of the \co, hereafter referred to as CO, molecule) components, as stars form in galaxies from the giant molecular clouds (GMCs) that arise out of the cold ISM. The molecular component of the cold ISM, which is found to be more centrally concentrated in spiral disks, tends to more closely trace the sites of recent star-formation activity than the \hi gas, which is more extended and loosely-bound to the galactic disk \citep[see][and references therein]{youngscoville1991}. 

Studies of the total cold ISM (molecular and atomic) in galaxies have historically been relegated to the very local Universe (at distances $\lesssim$100 h$^{-1}$Mpc), with most of the environmental studies focusing on the Virgo cluster and Coma supercluster \citep[e.g., ][]{haynes1984, giovanellihaynes1985, gavazzi1987, kenneyyoung1989, casoli1996, boselli1997, gavazzi2006, pappalardo2012, scott2013, boselli2014}. These studies generally found strong evidence for the \hi gas being more readily stripped in the cluster environment than the molecular component, compared to field galaxies. \citet{fabello2011} showed, by \hi stacking on the Arecibo Legacy Fast ALFA survey \citep[ALFALFA;][]{giovanelli2005}, that galaxies at $z\leq0.06$ exhibit distinct atomic gas deficiencies in environments of higher local density. The COSMOS HI Large Extragalactic Survey \citep[CHILES;][]{fernandez2013} is surveying a portion of the COSMOS \citep{scoville2007} field with the Jansky Very Large Array, and with sufficient sensitivity to detect atomic gas in galaxies out to $z\sim0.5$. In recent years, observations of the cold molecular ISM in galaxies have begun extending to higher redshifts \citep[e.g.,][]{daddi2010, krips2012, magdis2012, aravena2012, bauermeister2013, combes2013, hodge2013, carilliwalter2013}, and to probe higher-density environments, including luminous infrared galaxies (LIRGs) in the outskirts of intermediate-redshift clusters \citep{geach2009, geach2011, jablonka2013}, but \hi observations are much more scarce at intermediate redshifts. The \coshort Legacy Database for the GALEX Aricebo SDSS Survey \citep[COLD GASS;][]{saintonge2011a} has recently combined molecular and atomic gas measurements, as well as measures of star-formation activity and other galaxy properties, for a mass-selected sample of 366 galaxies with stellar masses $M_*\gtrsim10^{10}\msun$ and redshifts $z\leq0.05$. 

In recent years, simulations of the gas content in galaxies have begun to predict, using semi-analytic \citep[e.g.][]{obreschkow2009, lagos2011, lagos2014, popping2014, popping2015} or hydrodynamical \citep[e.g.][]{dave2012, dave2013, rafieferantsoa2014} prescriptions, how gas evolves in galaxies over time. One of the most important unanswered questions is how the atomic gas content of galaxies evolves with time relative to the molecular gas. \citet{popping2014}, and others, have shown that changes in model assumptions can yield dramatically different evolution of the molecular-to-atomic gas ratios in galaxies, and the lingering uncertainties regarding the abundances of these cold gas components significantly obfuscate our theoretical understanding of galaxy evolution. These model uncertainties underscore a need for observations of the total cold gas content of the ISM that extend beyond the local Universe, and which sample a range of galaxy environments.

To address the observational need for measurements of atomic gas in galaxies at intermediate redshifts, and at higher-density environments, we have carried out an unprecedented study of the atomic gas, stellar populations, morphology, and star-formation activities of two galaxy clusters at $z\sim0.2$ and their surrounding large scale structure. Our project, the Blind Ultra-Deep \hi Environmental Survey \citep[BUDHIES;][]{verheijen2007, verheijen2010, jaffe2012, jaffe2013, jaffe2015}, consists of multi-wavelength observations covering an area of $1\times1$ deg$^2$ ($\sim$100 Mpc$^2$) around the two clusters Abell 2192 ($z=0.188$, RA=16:26:37.1, Dec=+42:40:20) and Abell 963 ($z=0.206$, RA=10:17:13.9, Dec=+39:01:31). The \hi data allow us to sensitively probe the effects of galaxy transformation (e.g., ram-pressure stripping, starvation, harassment, and mergers) on atomic gas in galaxies in a range of environments. These two clusters were chosen for their contrasting dynamical states; A963 is extremely rich, X-ray luminous, and fairly dynamically relaxed, while A2192 is less massive, less X-ray luminous, and is less relaxed \citep{jaffe2013}. Both clusters have been shown to contain significant sub-structure in our spectroscopic studies \citep{jaffe2013}. All together, our study provides an unprecedented look at the evolutionary state of galaxies in a large dynamic range of environments, and at a redshift where the Butcher-Oemler effect \citep{butcheroemler1984} first presents a strong increase in galaxy activity at high densities.

To fill in the missing pieces of the cold gas puzzle in these two clusters, we have begun a pilot study of the molecular gas content of BUDHIES galaxies with the Redshift Search Receiver (RSR) on the Large Millimeter Telescope {\it Alfonso Serrano} (LMT) in Mexico. Our pilot study, which we are calling \textbf{CO} \textbf{O}bservations with the \textbf{L}MT of BUDHIES (COOL BUDHIES), has measured the CO emission line, or placed upper limits on the emission line, for a sample of 23 galaxies selected from the BUDHIES fields. Our sample, which is comprised of half \hi-selected galaxies and half \hi-undetected but selected via detections in the infrared with the Multiband Imaging Photometer for {\it Spitzer} \citep[MIPS;][]{rieke2004}, consists of targets with stellar masses ${\rm M}_*\geq10^{10}\msun$ and spectroscopic redshifts from the optical and/or \hi.

Section \ref{sample} describes our sample and our existing BUDHIES data (\ref{data_budhies}), and our new LMT \coshort observations are described in Section \ref{data_co}. Section \ref{datareduct} describes our procedures for reduction and analysis of the new \coshort spectra. In Section \ref{cg} we describe our primary reference sample, and in Section \ref{results} we present our molecular gas masses (\ref{mol_gas}), and compare the gas content of our target sample to our local reference sample. Section \ref{mol_vs_atomic} compares the molecular and atomic gas masses in our sample, and to our reference sample, and Section \ref{gas_and_environ} examines the gas content related to environment. We discuss the implications of our results, further interpretation, and highlight the next steps of the COOL BUDHIES project in Section \ref{conclusions}. Throughout this paper we use cosmological parameters $\Omega_{\Lambda}=0.70$, $\Omega_{\rm M}=0.30$, and $H_0=70 \;{\rm km\; s}^{-1} \;{\rm Mpc}^{-1}$, where pertinent cosmological quantities have been calculated using the online Cosmology Calculator of E.~L.~Wright \citep{wright2006}. We also assume a Kroupa IMF \citep{kroupa2001}.

\section{Sample and Data}\label{sample}

\subsection{BUDHIES Sample}\label{data_budhies}

The foundation of the BUDHIES project is ultra-deep \hi mapping with the Westerbork Synthesis Radio Telescope (WSRT), with 78$\times$12hr on A2192 and 117$\times$12hr on A963, to a 4$\sigma$ detection threshold of $2\times10^{9}\msun$. The details of the \hi observations, the data reduction, and catalogue generation can be found in \citet{verheijen2007, verheijen2010}. Our WSRT survey revealed $\sim$160 \hi detections spanning redshifts of $0.164\leq z\leq 0.224$. To supplement these data we have obtained imaging with the {\it Galaxy Evolution Explorer} \citep[GALEX;][]{martin2005} in NUV and FUV, B- and R-band with the INT on La Palma, United Kingdom Infrared Telescope (UKIRT) J- (for A963), H-, and K-band, and {\it Spitzer} Infrared Array Camera \citep[IRAC;][]{fazio2004} and MIPS. To more fully sample the optical-to-NIR part of the spectrum, we also obtained data from the Sloan Digital Sky Survey \citep[SDSS;][]{york2000}. For the SDSS, we generated mosaics in $u^\prime$$g^\prime$$r^\prime$$i^\prime$$z^\prime$ with the online Montage Image Mosaic Service\footnote{\url{http://hachi.ipac.caltech.edu:8080/montage/}}, which produces science-grade mosaics by co-adding SDSS frames over an area of up to 1 square degree. We also have spectroscopic redshifts for over 2000 galaxies in these two clusters, which come from a combination of spectra taken at the WHT in La Palma, the SDSS, the Wisconsin Indiana Yale NOAO (WIYN) Telescope, and MMT/Hectospec observations from \citet{hwang2014}, and from the Local Cluster Substructure Survey (LoCuSS) team (private comm.). Details can be found in \citet{jaffe2013} and in Jaff\'{e} et al. (in prep).

The {\it Spitzer} IRAC and MIPS data (PI: A. Chung) were reduced using the IDL pipeline of R. Gutermuth \citep{gutermuth2009}, and we use the MIPS [24] data to estimate the total infrared luminosity $L_{\rm IR}$ following the calibration of \citet{rieke2009}. The UKIRT Near-IR data (PI: G. Morrison; JHK for A963 and HK for A2192) were processed by the JAC pipeline, and co-added mosaics were produced using the complementary Astromatic\footnote{\url{http://www.astromatic.net}} tools Source Extractor \citep{bertinarnouts1996}, SWarp \citep{bertin2002}, and SCAMP \citep{bertin2006}. The {\it GALEX} FUV and NUV photometry (PI: J. H. van Gorkom) come from the reduced data products available from the Mikulski Archive for Space Telescopes\footnote{\url{http://galex.stsci.edu/}} (MAST), which provides calibrated photometry catalogues for our maps of these two clusters.

Obtaining accurate stellar masses for our sample of galaxies can only be done with properly-calibrated photometry and band-merged catalogues. We de-redden the photometry using the foreground galactic extinction values from \citet{schlegel1998}, assuming $R_{\rm V}=3.1$. The INT R-band image, being the deepest and highest-resolution map of these clusters, forms the basis of our photometric catalogue. To make our band-merged catalogues, we first matched the astrometry of all our other images to the INT R-band frame, correcting for sub-arcsecond offsets that we measured using bright (but un-saturated) point sources detected using Source Extractor in each frame. In the process of checking the astrometry, we found a common occurrence of a systematic offset in Declination of $\sim0.3\arcsec$ for all of our frames compared to the INT B- and R-band frames, and so we adjusted the INT astrometry to be in better agreement with the median systematic offsets in Declination as well. After getting each frame onto a common astrometric solution, we measure photometry with Source Extractor using Kron elliptical apertures for all bands from SDSS $u^\prime$ through IRAC [4.5]. We measure aperture corrections in each of these frames by comparing the elliptical-aperture photometry, for isolated sources, with the photometry measured from much larger circular apertures, obtaining corrections of approximately 0.05-0.10 mag in each band. Finally, the individual catalogues are merged with the INT R-band source list to produce a final catalogue. The IRAC [5.8] and [8.0] photometry are excluded from our catalogues, as we found them to generally lack sufficient sensitivity. After merging the SDSS $u^\prime$ through IRAC [4.5] bands, we match the {\it GALEX} FUV \& NUV catalogue as well as the MIPS [24] catalogue to the final band-merged catalogue.

After band-merging, we perform spectral energy distribution (SED) fitting using the Fortran-based code MAGPHYS\footnote{\url{http://www.iap.fr/magphys/magphys/MAGPHYS.html}} \citep{dacunha2008, dacunha2015}. SED fitting is restricted to only those galaxies having spectroscopic redshifts (either from optical or \hi), keeping the redshift fixed and finding the best-fitting SED from the \citet{bruzualcharlot2003} population synthesis models. The MAGPHYS code is built with a Bayesian framework, and it marginalizes over a number of parameters affecting the stellar light (e.g., metallicity and dust extinction) and it also can simultaneously find the best-fitting dust emission in the infrared, while maintaining energy balance between the absorbed UV-optical light and the re-emitted infrared (via Polycyclic Aromatic Hydrocarbons in addition to warm and cold dust components). Since we are only concerned with the stellar component of the SED in our present analysis, and we only fit SEDs using the {\it GALEX} through IRAC [4.5], we ignore any of the dust information returned by MAGPHYS, and use just the total stellar mass (converted to a Kroupa IMF). To estimate the typical $1\sigma$ dispersion of our stellar mass estimates, we exploit the fact that MAGPHYS returns a full probability distribution function (PDF) of the stellar mass. We stack on all of the stellar mass PDFs, centered on the maximum likelihood stellar mass of each, for all galaxies having a stellar mass $M_*\geq10^{10}\msun$. The mean stacked PDF has a standard deviation of $\simeq0.08$ dex, which we conservatively round up to 0.1 dex to help account for additional systematic uncertainties affecting our stellar mass estimates.

To verify that we have obtained reasonable mass estimates, we compare our stellar masses to those calculated using independent calibrations in the optical and near-to-mid-infrared. Our comparison optically-derived stellar masses are from  \citet{zibetti2009}, using our INT $B-$ and $R-$band rest frame photometry with:

\begin{equation}
\frac{M_*}{\msun} = L_R\times10^{-1.200 + 1.066({\rm mag}_B-{\rm mag}_R)}+10^{0.04},
\end{equation}

where $L_R$ is the $R$-band luminosity (in $\lsun$) and the $10^{0.04}$ term converts the IMF to Kroupa. Our other comparison stellar mass calibration comes from \citet{eskew2012} using IRAC [3.6] \& [4.5]. We estimate stellar masses, similarly as we have in \citet{cybulski2014}, as: 

\begin{equation}
log\left(\frac{M_*}{\msun}\right) = log(0.69\times10^{5.65})f_{I1}^{2.85}f_{I2}^{-1.85}(D_L/0.05)^2
\end{equation}

where $f_{I1}$ and $f_{I2}$ are the rest-frame fluxes in [3.6] and [4.5], respectively, in Jy, $D_L$ is the luminosity distance in Mpc, and the mass is also in a Kroupa IMF. 

Using the $\sim2000$ galaxies in these two fields that have spectroscopic redshifts, detections in the optical and IRAC [3.6] and [4.5] bands, and a stellar mass range of $10^8 \leq M_* \leq 10^{12}\msun$, we compare the \citet{zibetti2009} optical stellar masses and the \citet{eskew2012} IRAC stellar masses to those of MAGPHYS. We find a strong linear correlation (with a Pearson correlation coefficient of 0.91 and 0.80 for the optical and IRAC stellar masses, respectively), a median stellar mass agreement of within 20 per-cent, and a dispersion corresponding to 0.25 dex (for optical) and 0.32 dex (for IRAC). Hereafter, our stellar masses come from the MAGPHYS code.

Our targets were selected from our band-merged catalogue for having:

\begin{enumerate}
\item $M_* \geq 10^{10} \msun$ 
\item spectroscopic redshifts, with $\mid z_{\rm spec} - z_{\rm cl}\mid \leq 0.05(1+z_{\rm cl})$, where $z_{\rm cl}$ is the redshift of the cluster
\item \textbf{either} a detection in \hi, \textbf{or} no detection in \hi but a detection in MIPS [24].
\end{enumerate}

The full sample of galaxies matching these selection criteria is over 150, but for the purposes of our pilot study we must restrict our observations to a small subset. Therefore, our sample consists of roughly half galaxies that are \hi selected, with no regard for their MIPS [24] flux, and half that are undetected in \hi but are MIPS [24] selected. Note, however, that our redshift window for target selection for \coshort observations doesn't overlap exactly with the redshift window for our \hi detections with the WSRT ($0.1646\leq z_{\rm HI}\;\leq 0.2241$). As a result, the eleven galaxies lacking \hi data in our sample are comprised of six which are in the volume mapped by the WSRT, and have upper limits on their \hi masses, and five which are outside of that volume (for which we have no data). Hereafter, we define galaxies as cluster members if they have projected separations of $R_{\rm proj}\leq3R_{200}$ from the cluster center and line-of-sight velocities within three times the velocity dispersion of the cluster.

\subsection{New LMT Observations}\label{data_co}

The LMT is a 50-m radio telescope located on Volc\'{a}n Sierra Negra in Mexico, at an elevation of 4600 meters \citep{hughes2010}. For the Early Science campaigns at the LMT, the inner 32.5 meters of the primary dish is illuminated by the receiver optics. During the observing seasons, the median opacity at the site at 225 GHz is $\tau=0.1$. The pointing RMS is 3$\arcsec$ over the entire sky, but is reduced to 1-2$\arcsec$ for targets located within $\sim10\deg$ of known sources.

We observed our targets with the RSR between 13 March and 29 April of 2014 as part of the Early Science 2 (ES2) season at the LMT. The RSR has a novel design, with a monolithic microwave integrated circuit (MMIC) system, that receives signals over four pixels simultaneously covering a frequency range of $73-111$ GHz, sampled at 31 MHz (corresponding to $\sim100 {\rm km}\; {\rm s^{-1}}\;$ at 90 GHz). The RSR has a beam full width half maximum (FWHM) that is frequency-dependent, but for our targets it is $\simeq23\arcsec$. The RSR beam FWHM is very well-matched to the angular sizes of the optical disks of our target galaxies, whose median $R_{90}$, derived from our INT R-band mosaic, is 11.6$\arcsec$ (see the postage-stamp images of our targets in Appendix \ref{allspectra} for a comparison of the optical disks to the RSR beam). The RSR system has been optimised to provide great stability in spectral baseline over the entire frequency range being sampled. The RSR was designed to operate on the LMT, but it has previously been commissioned on the Five College Radio Astronomy Observatory (FCRAO) 14-m telescope \citep[e.g.,][]{chung2009, snell2011}, and was also used recently with LMT observations in \citet{kirkpatrick2014} and \citet{zavala2015}. For a technical description of the RSR system, see \citet{erickson2007}. Our observations were taken with a system temperature ranging from $87-113$K, and our targets were observed for about 1 hr each (see Table \ref{tab:table_rsr_obs} for specific integration times) with typical rms noise of $\sim0.190{\rm mK}$.

\subsubsection{Data Reduction and Analysis}\label{datareduct}

We reduced the spectra using DREAMPY (Data REduction and Analysis Methods in PYthon), a software package written by G. Narayanan specifically to reduce and analyse RSR spectra. The RSR produces four separate spectra for each observation; prior to co-adding them, the four spectra are individually calibrated and visually checked for any known instrumental artifacts that occasionally arise. Any portion of the spectrum found to exhibit those artifacts is flagged for removal.

\begin{figure}
	\includegraphics[width=3.4in]{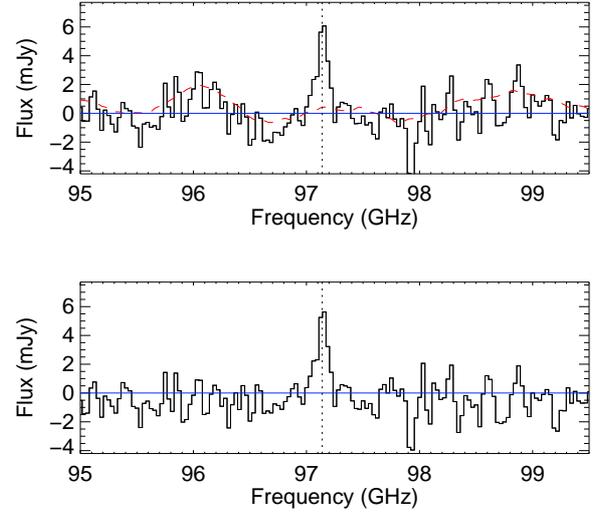}
	\caption{A portion of the spectrum of our target J162523.6+422740, centered on the CO line. The top panel shows the spectrum after reducing it with DREAMPY, without applying our Savitzky-Golay filter. The polynomial fit to the spectrum is denoted by the red dashed line. The bottom panel is the spectrum after filtering. The vertical dotted line in both panels shows the expected central frequency of the line, based on the prior redshift information.}
	\label{fig:example_baseline}
\end{figure}

\begin{figure}
	\includegraphics[width=3.4in]{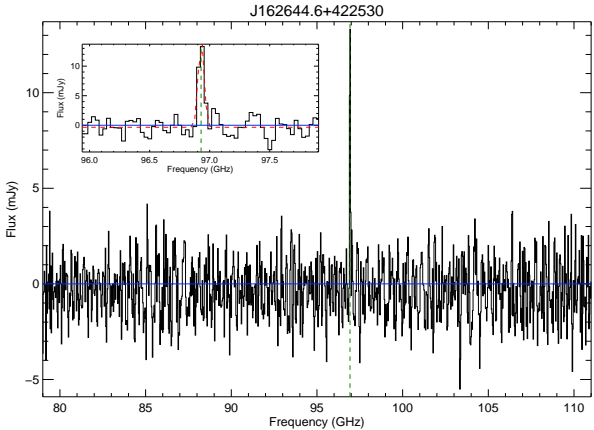}
	\caption{CO spectrum of one of our targets (J162644.6+422530), after filtering, with a strong detection of the \colong transition line. The vertical dashed green line indicates the frequency where we expect to detect the line, based on the \hi redshift. The full spectrum is seen in the main figure and a zoomed-in view, centered on the frequency of the CO line, is in the inset. The dashed red line in the inset denotes the Gaussian fit to the CO line.}
	\label{fig:example_fullspec}
\end{figure}

After co-adding the spectra, we analyse them using a custom IDL code that fits the line with a Gaussian using a Markov Chain Monte Carlo (MCMC) approach to robustly determine the parameters of the line (amplitude, central frequency, standard deviation, and D.C. offset) and their statistical errors. We begin by searching for an initial Gaussian fit in the spectrum to a line having positive amplitude and a central frequency within $\pm$0.08 GHz of the expected \coshort frequency (corresponding to a velocity range of $\pm \sim250$km s$^{-1}$), based on the prior optical/\hi redshift for each galaxy. Then we subtract off that best-fitted Gaussian from the spectrum and apply a Savitzky-Golay filter \citep{savitzkygolay1964} to the full spectrum that remains, to reduce the low-frequency signal. The Savitzky-Golay filter we use is a ``rolling'' order-two polynomial fit to the spectrum with a width of 1 GHz. Note that the width of our Savitzky-Golay filter is significantly greater than the width of any astrophysical lines in our spectra. This filtering technique has been employed in many prior spectroscopic studies \citep[e.g.,][]{faran2014, stroe2015, wang2015}. After computing the polynomial filter on the line-subtracted spectrum, we apply that filter to the original spectrum, and then we fit a final Gaussian to the CO emission line in the filtered spectrum. A demonstration of the filtering applied to one of our spectra is shown in Figure \ref{fig:example_baseline}. In Figure \ref{fig:example_fullspec} we show the full spectrum of one of our targets, as well as a zoomed-in view of the portion of the spectrum near the identified CO line.

We convert the spectra from units of modified antenna temperature ${\rm T}^*_{\rm A}$ to flux density by multiplying by the telescope gain of $7 {\rm \;Jy \;K}^{-1}$ (F.~P.~Schloerb, private comm.). And we also convert the spectra from units of frequency to velocity, centered on the $\nu_{\rm CO}$ that we fit in the spectrum, and account for distortions when translating between frequency intervals and velocities at each galaxy's redshift following \citet{gordon1992}. Then we obtain the total line flux by integrating the best-fitted Gaussian over $\pm2\sigma$ centered on the velocity of the line. Table \ref{tab:table_rsr_obs} presents the basic results of the \coshort observations, including the integrated line fluxes, full width half maximum (FWHM), and derived \coshort redshifts. Appendix \ref{allspectra} presents all of our \coshort spectra, separated into cluster members and non-members.

\begin{figure}
	\includegraphics[width=3.5in]{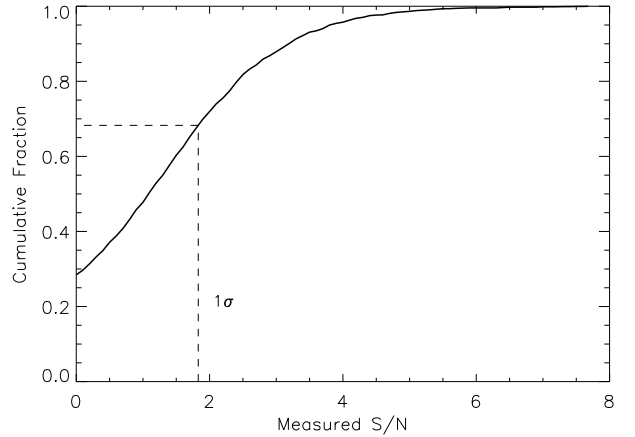}
	\caption{Cumulative distribution of signal-to-noise measured using our MCMC code for 2000 randomly-selected frequencies in our RSR spectra (avoiding parts of the spectrum where we expect a \coshort line). The dashed line indicates the signal-to-noise values corresponding to 1 times the standard deviation.}
	\label{fig:cumul_line_test}
\end{figure}

Given that our technique for identifying molecular emission lines assumes a particular central frequency, based on prior redshift information, and then searches for the best-fitting Gaussian near that frequency, we want to be sure that we understand the statistical significance of our signal-to-noise measurements. To better understand our statistics, we select 2000 random combinations of frequency and target number (avoiding the parts of the spectra where we know or expect a \coshort line to be located) and run our filtering and line detection algorithm on each of our random selections, which presumably only consist of noise. Then we use the cumulative distribution of the signal-to-noise values recovered in these random trials to determine at which of our measured signal-to-noise values do we truly find the standard deviation. ${\rm S/N}_{1\sigma}$ corresponds to the signal-to-noise where our cumulative distribution reaches 68.269\%, and ${\rm S/N}_{2\sigma}$ is when the cumulative distribution hits 95.45\%. These tests reveal that ${\rm S/N}_{1\sigma}=1.8$, and ${\rm S/N}_{2\sigma}=3.6$, as shown in Figure \ref{fig:cumul_line_test}. Based on these tests, we decide to count any detections with $2.0<{\rm S/N}<3.6$ as ``marginal'' and only consider a ${\rm S/N}\geq3.6$ to be a reliable detection. We do consider the integrated line flux, and estimated molecular gas mass, for marginal detections, but we do not derive any other parameters (e.g., ${\rm CO}$ redshifts) for them.

\begin{table*}
 \centering
  \caption{\coshort observations of our target galaxies, separated into members of the two clusters and foreground/background galaxies around the clusters. Column 2 gives the redshift of the target, based on prior optical or \hi observations. Column 3 gives the integration time. Column 4 has the RMS of the RSR spectrum, measured over a frequency range of $\pm$1 GHz centered on the \coshort line (excluding the line itself). Column 5 has the integrated line flux, and Column 6 gives the central frequency of the \coshort line. Column 7 has the FWHM of the line, and Column 8 the redshift derived from the \coshort line. Note that we only give the latter of the derived quantities for the cases where we have a reliable detection (${\rm S/N}\geq3.6$)} \label{table_rsr_obs}
  \begin{tabular}{@{}llllllll@{}}
  \hline
  Designation  &  z$_{\rm opt/HI}$ & t$_{\rm int}$ & rms   & S$_{\rm CO}\Delta$V     &  $\nu_{\rm CO}$ & $\Delta$V & z$_{\rm CO}$ \\
               &     &  (hr)&  (mK) & (Jy km s$^{-1}$)  &  (GHz)  &   (km s$^{-1}$)  &       \\
 \hline
\textbf{A2192 Galaxies}
   &           &     &           &                    &         &  \\
 J162523.6+422740 & 0.187 & 2.0 & 0.154 & 1.995 $\pm$ 0.294 & 97.1387 $\pm$ 0.0110 & 429 $\pm$ 52 & 0.18667 $\pm$ 0.00061 \\ 
 J162644.6+422530 & 0.189 & 1.0 & 0.217 & 2.387 $\pm$ 0.293 & 96.9290 $\pm$ 0.0022 & 176 $\pm$ 12 & 0.18923 $\pm$ 0.00012 \\   
 J162528.4+424708 & 0.189 & 0.9 & 0.248 & $<$1.006 & ... & ... & ... \\   
 J162508.6+423400 & 0.190 & 1.0 & 0.218 & 1.326 $\pm$ 0.466 & ... & ... & ... \\   
\hline
\textbf{A2192 FG/BG Galaxies}
   &             &     &           &                    &        &   \\
 J162555.2+425747 & 0.134 & 1.0 & 0.246 & 4.708 $\pm$ 0.550 & 101.6751 $\pm$ 0.0128 & 598 $\pm$ 49 & 0.13372 $\pm$ 0.00094 \\    
 J162612.9+425242 & 0.146 & 1.0 & 0.196 & 3.329 $\pm$ 0.256 & 100.6082 $\pm$ 0.0019 & 161 $\pm$ 11 & 0.14574 $\pm$ 0.00013 \\   
 J162558.0+425320 & 0.169 & 1.9 & 0.175 & $<$0.856 & ... & ... & ... \\    
 J162710.8+422754 & 0.173 & 1.0 & 1.485 & $<$4.202 & ... & ... & ... \\  
 J162721.0+424951 & 0.220 & 2.1 & 0.137 & 0.949 $\pm$ 0.268 & ... & ... & ... \\   
 J162613.4+423304 & 0.224 & 1.0 & 0.216 & $<$0.637 & ... & ... & ... \\  
 J162717.7+430309 & 0.228 & 1.1 & 0.211 & 0.714 $\pm$ 0.294 & ... & ... & ... \\   
 J162830.3+425120 & 0.228 & 1.1 & 0.195 & 1.392 $\pm$ 0.386 & 93.7779 $\pm$ 0.0118 & 559 $\pm$ 150 & 0.22919 $\pm$ 0.00055 \\   
\hline
\textbf{A963 Galaxies}
   &               &     &           &                    &       &    \\
 J101703.5+384157 & 0.201 & 1.0 & 0.224 & $<$0.919 & ... & ... & ... \\  
 J101727.7+384628 & 0.201 & 1.2 & 0.230 & 0.852 $\pm$ 0.314 & ... & ... & ... \\   
 J101705.5+384925 & 0.204 & 1.0 & 0.237 & $<$1.537 & ... & ... & ... \\  
 J101540.2+384913 & 0.204 & 1.0 & 0.325 & 1.292 $\pm$ 0.445 & ... & ... & ... \\  
 J101730.0+385831 & 0.204 & 1.0 & 0.186 & 1.324 $\pm$ 0.313 & 95.7115 $\pm$ 0.0144 & 340 $\pm$ 50 & 0.20436 $\pm$ 0.00073 \\   
 J101803.6+384120 & 0.205 & 1.1 & 0.141 & $<$0.411 & ... & ... & ... \\   
 J101611.1+384924 & 0.207 & 1.0 & 0.208 & 1.640 $\pm$ 0.286 & 95.5613 $\pm$ 0.0075 & 203 $\pm$ 38 & 0.20625 $\pm$ 0.00038 \\ 
 J101618.0+390613 & 0.208 & 1.0 & 0.253 & 1.771 $\pm$ 0.348 & 95.3854 $\pm$ 0.0052 & 198 $\pm$ 23 & 0.20848 $\pm$ 0.00026 \\   
\hline
\textbf{A963 FG/BG Galaxies}
   &              &     &           &                    &        &   \\
 J101856.7+390158 & 0.161 & 1.1 & 0.302 & 2.166 $\pm$ 0.690 & ... & ... & ... \\   
 J101712.2+390559 & 0.165 & 1.2 & 0.260 & $<$0.730 & ... & ... & ... \\    
 J101624.0+385840 & 0.169 & 2.0 & 0.163 & 2.146 $\pm$ 0.265 & 98.6013 $\pm$ 0.0067 & 319 $\pm$ 26 & 0.16906 $\pm$ 0.00040 \\    
\hline
\label{tab:table_rsr_obs}
\end{tabular}
\end{table*}

\subsection{Reference Sample: COLD GASS}\label{cg}

For any pilot study such as ours, it is extremely important to place our results into context with previous studies that have examined similar astrophysical quantities. The natural choice for a reference sample to our current study is the CO Legacy Database for the {\it GALEX} Arecibo SDSS Survey \citep[COLD GASS;][]{saintonge2011a}. COLD GASS is an IRAM 30-m legacy survey that targeted about 350 nearby ($z\leq0.05$) galaxies with stellar masses $M_*\geq10^{10}\msun$. The COLD GASS sample is mass-selected from the parent GASS survey \citep{catinella2010}, which consists of \hi observations with Arecibo for nearby massive galaxies selected from the SDSS and {\it GALEX}. We obtained the COLD GASS data products from their third public data release on their project website\footnote{\url{http://www.mpa-garching.mpg.de/COLD\_GASS/}}. The COLD GASS data set has also been used as a reference sample in \citet{jablonka2013} and in \citet{lee2014}. 

To provide a better comparison with our own study, we supplement the available data products for COLD GASS with photometry from the Wide-Field Infrared Survey Explorer \citep[WISE,][]{wright2010}, which has mapped the whole sky in 3.4, 4.6, 12, and 22 $\mu{\rm m}$. As in \citet{cybulski2014}, we found matches to WISE by searching in the AllWISE catalogue for counterparts within a $5\arcsec$ search radius centered on the SDSS galaxy positions of the COLD GASS sample galaxies. We use the WISE [22] photometry to estimate $L_{\rm IR}$ and $SFR_{\rm IR}$, also using the calibration of \citet{rieke2009}. 

\subsubsection{Aperture corrections and beam contamination}

One key benefit of our targets being at higher redshift than those of the COLD GASS sample is that the beam for our \coshort observations more completely covers the disks of our galaxies than for our lower-z reference sample. Consequently, we can confidently measure the full extent of the \coshort emission in our targets without concern for missing any appreciable flux. \citet{saintonge2011a} showed that for the COLD GASS sample, with a beam approximately the same size as that of the RSR (22$\arcsec$), they require a range of aperture corrections for their \coshort fluxes from $\sim20-50$ per-cent, depending on the angular size of the galaxy. If we apply their aperture correction formula to our targets, using the measured optical sizes of our targets, we would require corrections of less than 2 per-cent. Given that our measurement uncertainties are significantly greater than this correction factor, we opt not to apply these aperture corrections for our sample. The COLD GASS catalogues that we compare our sample with had aperture corrections applied to their \coshort measurements. For a comparison of the angular sizes of our targets in contrast with those of the COLD GASS sample, see the figures in Appendix \ref{allspectra}.

However, a potential drawback of the relative size of our aperture being greater, compared to the angular extent of our target galaxies, is the risk of contamination from nearby galaxies in our beam. This is a particularly significant concern when observing targets in crowded fields, like in galaxy clusters such as ours. Although we occasionally find additional optical detections in our maps within the RSR beam, we do not typically encounter multiple targets that are bright in the infrared in our beam (the one exception is J162721.0+424951, although in that case the dominant source of the infrared emission is our primary RSR target). Given that the strength of the CO line correlates strongly with infrared emission (see Figure \ref{fig:lir_vs_lco}), we use our {\it Spitzer} MIPS data to assess possible \coshort contamination, and also to correct for it.

\section{Results}\label{results}

\subsection{CO Luminosities and Molecular Gas Masses}\label{mol_gas}

\begin{figure*}
\centering
	\includegraphics[width=6.2in]{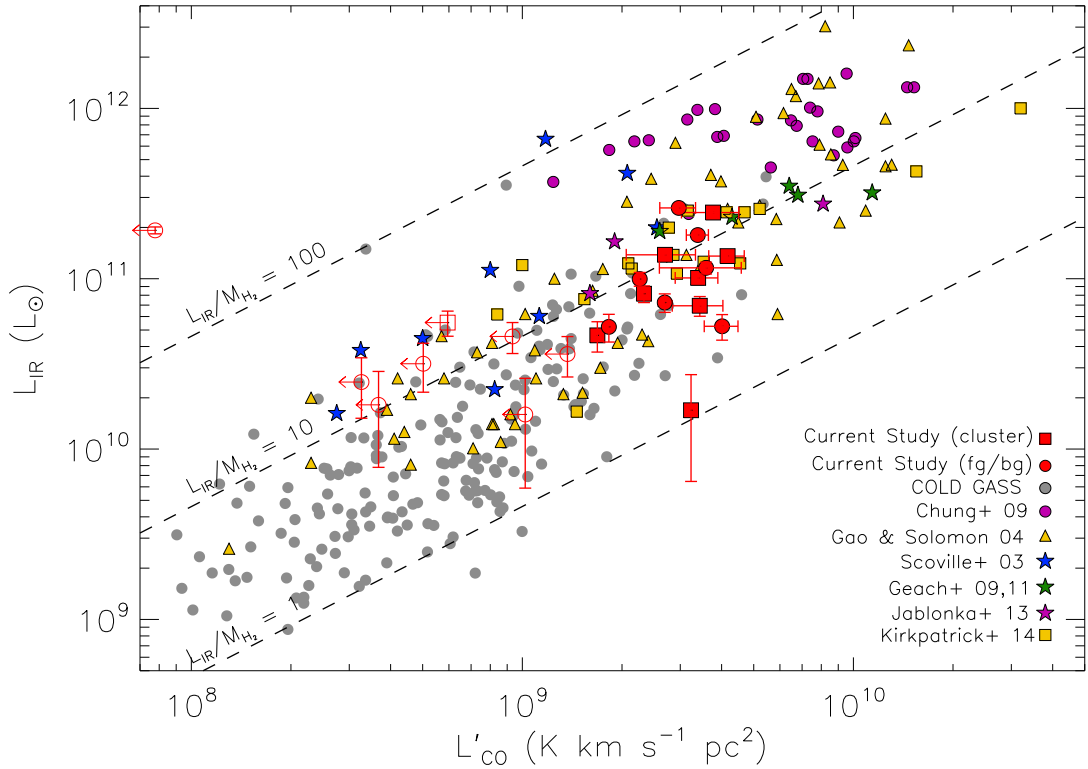}
	\caption{Infrared luminosity versus \coshort luminosity for our current sample (red squares are cluster members, and red circles are targets in the foreground or background of our clusters). Unfilled red squares and circles indicate non-detections. The dashed lines indicate constant values of the ratio of ${\rm L}_{\rm IR}\;$ to ${\rm M}_{\rm H_2}\;$ (both in solar units). We also compare our sample to a number of other studies collected in the literature. The grey circles are galaxies from COLD GASS \citep{saintonge2011a} with detections in \coshort and WISE [22]. The purple circles are nearby ULIRGs from \citet{chung2009}, observed with the RSR on the FCRAO 14-m telescope. Yellow triangles correspond to the sample of \citet{gaosolomon2004}. Blue stars are low-redshift QSOs from \citet{scoville2003}, and the green and blue squares indicate the intermediate-redshift cluster galaxies from \citet{geach2009, geach2011, jablonka2013}. The yellow squares are from \citet{kirkpatrick2014}. Note that the infrared luminosities being plotted for the \citet{kirkpatrick2014} points are not corrected to remove the contribution due to an active galactic nucleus (AGN), to remain consistent with the rest of the data being plotted.}
	\label{fig:lir_vs_lco}
\end{figure*}

We calculate the \coshort line luminosity by

\begin{equation}\label{co_lum}
\frac{L^\prime_{\rm CO}}{{\rm K}\; {\rm km}\; {\rm s}^{-1}\; {\rm pc}^2} = 3.25\times10^7 S_{\rm CO}\Delta {\rm V} \nu^{-2}_{\rm obs}\:D^2_L(1+z)^{-3},
\end{equation}

following \citet{solomonvandenbout2005}, where $\nu_{\rm obs}\;$ is the frequency of the line in GHz and $D_L$ is the luminosity distance in Mpc. In Figure \ref{fig:lir_vs_lco} we plot the infrared luminosities versus \coshort line luminosities for our target sample, compared to similar observations gathered from the literature \citep{scoville2003, gaosolomon2004, chung2009, geach2009, geach2011, jablonka2013, kirkpatrick2014}. Our galaxies that are detected in \coshort follow the established trends in the literature, and they mostly occupy an intermediate space between the less-infrared-luminous galaxies of COLD GASS and the ultra-luminous infrared galaxies (ULIRGs) from \citet{chung2009}. Furthermore, in Figure \ref{fig:lir_vs_lco} we see no apparent difference between the cluster members and the foreground/background galaxies in our sample. 

To obtain estimates of molecular gas masses, we assume a CO-to-$\htwo$ conversion factor of $\alpha_{\rm CO}=4.6 \msun ({\rm K}\; {\rm km}\; {\rm s}^{-1}\; {\rm pc}^2)^{-1}$, which is roughly the value observed in the Milky Way \citep{bolatto2013}, and implies the following conversion:

\begin{equation}
\frac{M_{H_2}}{\msun} = \frac{4.6L^\prime_{\rm CO}\;}{{\rm K}\; {\rm km}\; {\rm s}^{-1}\; {\rm pc}^2}.
\end{equation}

Table \ref{tab:table_results} gives the resulting \coshort line luminosities, infrared luminosities, and baryonic mass components (molecular gas, atomic gas, and stellar) for the galaxies in our sample. It is worthwhile to note that we unfortunately have only one galaxy in our sample that is not a cluster member and has detections in both molecular and atomic gas. This is partly due to small number statistics, but it is also a consequence of the target selection window in redshift space ($\mid z_{\rm spec} - z_{\rm cl}\mid \leq 0.05(1+z_{\rm cl})$) being a bit wider than the redshift window over which our WSRT mapping can detect galaxies in \hi ($0.1646\leq z_{\rm HI}\;\leq 0.2241$), as first mentioned in Section \ref{sample}. When we lack an \hi detection due to a non-detection in the \hi mapping, we indicate the upper limit on the \hi gas mass in Table \ref{tab:table_results}. However, when we lack an \hi detection because the target is outside of the redshift range of the \hi spectrum, we denote the \hi gas mass with ``...'' in Table \ref{tab:table_results} and we exclude these targets from any figures involving \hi gas mass hereafter.

\begin{table*}
 \centering
  \caption{Relevant luminosities and masses for our target galaxies. Column 2 gives the molecular gas line luminosity. Columns 3, 4 and 5 give the molecular gas, atomic gas, and stellar mass, respectively. Column 6 shows the infrared luminosity.} \label{tab:table_results}
  \begin{tabular}{@{}llllll@{}}
  \hline
  Designation  &  $L^\prime_{\rm CO}$  &  $M_{H_2}$ & $M_{HI}$ & $M_*$   &  log($L_{\rm IR}$)    \\
               &  (10$^9 {\rm K}\; {\rm km}\; {\rm s}^{-1}\; {\rm pc}^2)$ &  ($10^9\msun$)   &  ($10^9\msun$) &  ($10^{10}\msun$) & [$\lsun$]  \\
 \hline
\textbf{A2192 Galaxies}
   &             &                 &       &           &                             \\
 J162523.6+422740 & 3.40 $\pm$ 0.50 & 15.62 $\pm$ 2.30 & $<$2.00 & 1.77 & 11.00 \\ 
 J162644.6+422530 & 4.17 $\pm$ 0.51 & 19.18 $\pm$ 2.36 & 6.36 $\pm$ 0.53 & 6.00 & 11.13 \\ 
 J162528.4+424708 & $<$1.75 & $<$8.07 & 9.70 $\pm$ 0.46 & 1.49 & 10.74 \\ 
 J162508.6+423400 & 2.33 $\pm$ 0.82 & 10.72 $\pm$ 3.76 & $<$2.00 & 10.05 & 10.91 \\ 
\hline
\textbf{A2192 FG/BG Galaxies}
   &             &                 &       &           &                             \\
 J162555.2+425747 & 4.01 $\pm$ 0.47 & 18.46 $\pm$ 2.16 & ... & 3.75 & 10.72 \\   
 J162612.9+425242 & 3.39 $\pm$ 0.26 & 15.59 $\pm$ 1.20 & ... & 12.96 & 11.26 \\    
 J162558.0+425320 & $<$1.18 & $<$5.43 & 11.36 $\pm$ 0.71 & 3.16 & 10.39 \\   
 J162710.8+422754 & $<$6.06 & $<$27.88 & 6.53 $\pm$ 0.46 & 4.50 & 10.20 \\  
 J162721.0+424951 & 2.26 $\pm$ 0.64 & 10.41 $\pm$ 2.95 & 4.70 $\pm$ 0.45 & 4.29 & 11.00 \\ 
 J162613.4+423304 & $<$1.58 & $<$7.25 & 5.23 $\pm$ 0.49 & 1.81 & 10.50 \\ 
 J162717.7+430309 & 1.83 $\pm$ 0.75 & 8.40 $\pm$ 3.46 & ... & 2.93 & 10.72 \\ 
 J162830.3+425120 & 3.59 $\pm$ 0.99 & 16.51 $\pm$ 4.58 & ... & 10.48 & 11.06 \\  
\hline
\textbf{A963 Galaxies}
   &               &                 &       &           &                           \\
 J101703.5+384157 & $<$1.82 & $<$8.37 & $<$2.00 & 4.92 & 10.66 \\ 
 J101727.7+384628 & 1.69 $\pm$ 0.62 & 7.75 $\pm$ 2.86 & 10.00 $\pm$ 0.68 & 7.25 & 10.67 \\ 
 J101705.5+384925 & $<$3.13 & $<$14.41 & 8.44 $\pm$ 0.59 & 4.37 & 10.56 \\ 
 J101540.2+384913 & 2.62 $\pm$ 0.90 & 12.05 $\pm$ 4.15 & 9.34 $\pm$ 0.58 & 5.09 & 10.23 \\ 
 J101730.0+385831 & 2.70 $\pm$ 0.64 & 12.41 $\pm$ 2.93 & 3.51 $\pm$ 0.26 & 2.95 & 11.14 \\ 
 J101803.6+384120 & $<$0.85 & $<$3.91 & 16.80 $\pm$ 0.96 & 9.40 & 10.26 \\ 
 J101611.1+384924 & 3.44 $\pm$ 0.60 & 15.81 $\pm$ 2.75 & 13.54 $\pm$ 0.76 & 11.81 & 10.84 \\ 
 J101618.0+390613 & 3.77 $\pm$ 0.74 & 17.32 $\pm$ 3.41 & $<$2.00 & 6.37 & 11.39 \\ 
\hline
\textbf{A963 FG/BG Galaxies}
   &             &                 &       &           &                             \\
 J101856.7+390158 & 2.69 $\pm$ 0.86 & 12.39 $\pm$ 3.95 & ... & 12.23 & 10.86 \\ 
 J101712.2+390559 & $<$0.95 & $<$4.39 & $<$2.00 & 5.88 & 11.28 \\ 
 J101624.0+385840 & 2.97 $\pm$ 0.37 & 13.66 $\pm$ 1.69 & $<$2.00 & 9.73 & 11.42 \\ 
\hline
\end{tabular}
\end{table*}

\subsection{Molecular vs Atomic Gas Masses}\label{mol_vs_atomic}

\begin{figure*}
\centering
\begin{tabular}{cc}
\includegraphics[width=3.5in]{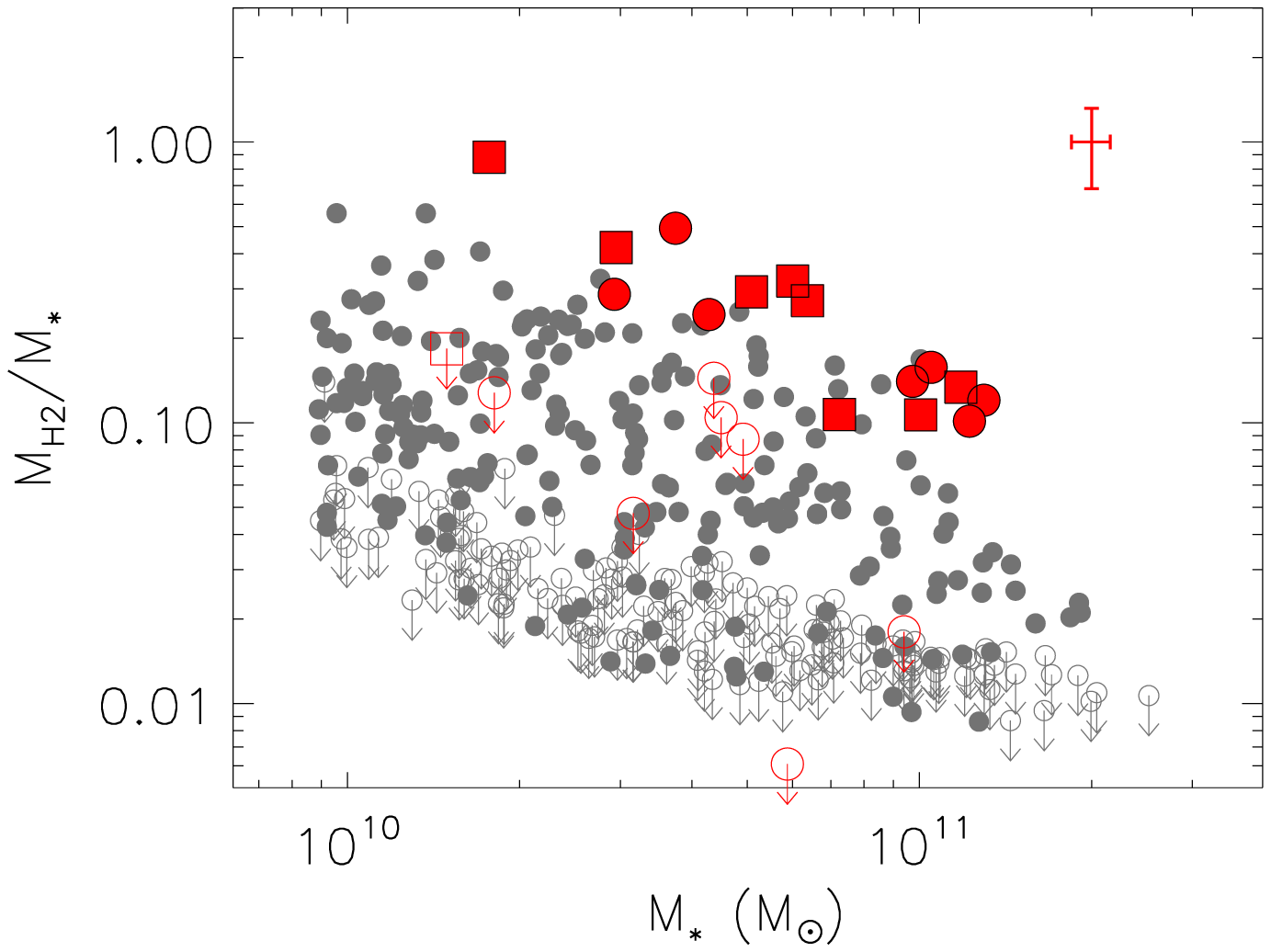}
\includegraphics[width=3.5in]{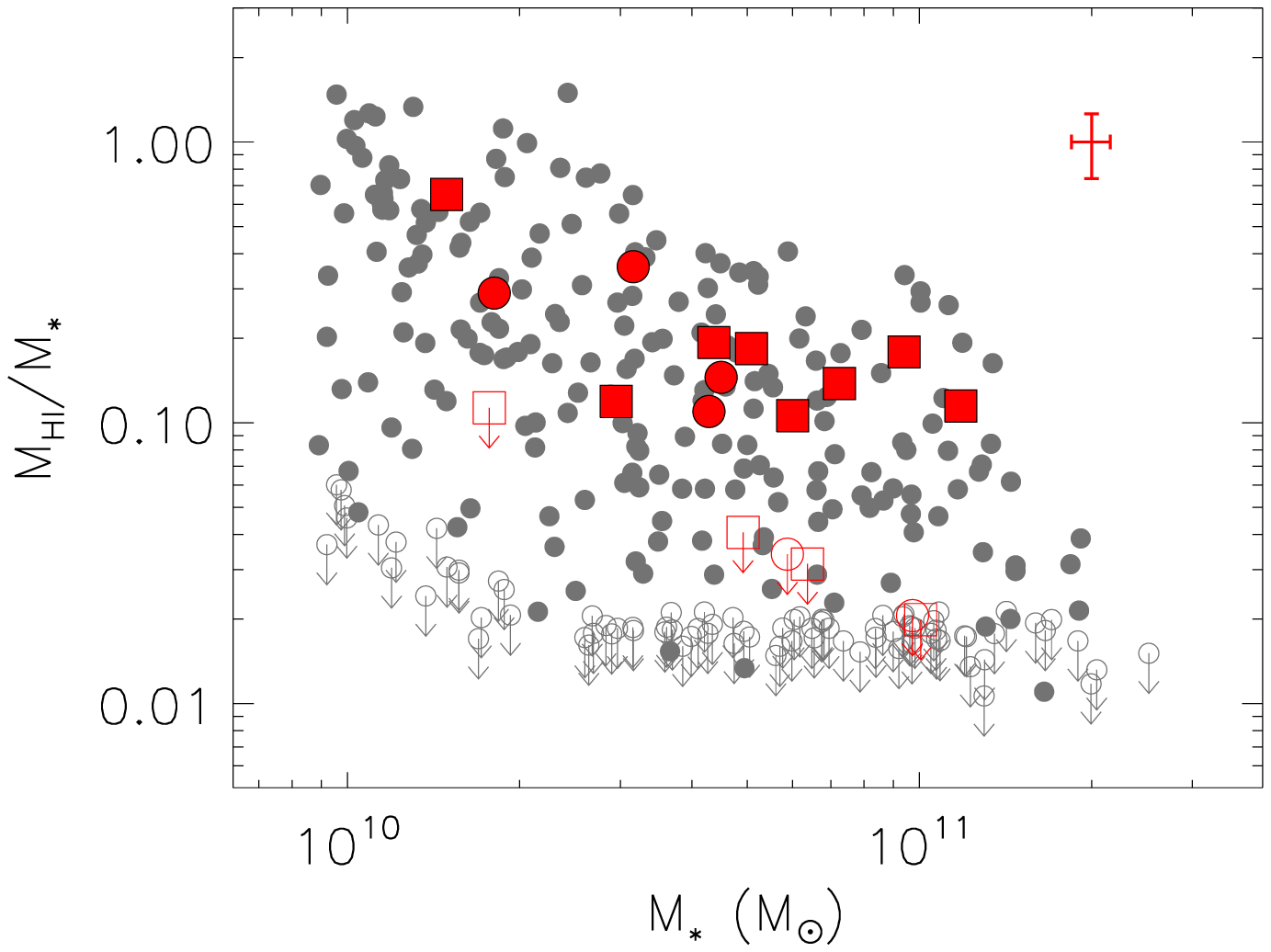}
\end{tabular}
\caption{(Left) Molecular gas masses, normalized by stellar mass, versus stellar mass for our target sample (in red) compared with the COLD GASS sample (grey). As in Figure \ref{fig:lir_vs_lco}, cluster members are squares and fg/bg galaxies are circles. Unfilled symbols indicate non-detections. (Right) Atomic gas masses, normalized by stellar mass, versus stellar mass for our target sample (in red) compared with the COLD GASS sample (grey).}
\label{fig:gas_masses}
\end{figure*}

In Figure \ref{fig:gas_masses} we plot a comparison of the molecular and atomic gas masses, normalized by stellar mass, between our targets and the COLD GASS sample. Note that the COLD GASS catalogs have molecular gas masses derived with a $\alpha_{\rm CO}$ of 4.35 (and 1.0 for the most infrared luminous galaxies), unlike our 4.6. In our comparisons, we have re-scaled the COLD GASS galaxies to match our adopted $\alpha_{\rm CO}$ factor throughout this work. It is notable that our detections in \coshort show molecular gas masses generally in excess of most of the COLD GASS sample, while our atomic gas masses show no such excess. However, given that our selection of targets is based in part on $L_{\rm IR}$, and that our threshold for detecting molecular gas is higher than with the COLD GASS sample, it is not surprising that our sample would produce molecular gas detections that are high relative to what is observed in the more local, not infrared-selected, sample of COLD GASS galaxies. Nevertheless, it is interesting that those same galaxies with high molecular gas masses, relative to the local reference sample, generally appear to have atomic gas masses that are more typical of the reference sample.

\begin{figure*}
\centering
\begin{tabular}{cc}
\includegraphics[width=3.5in]{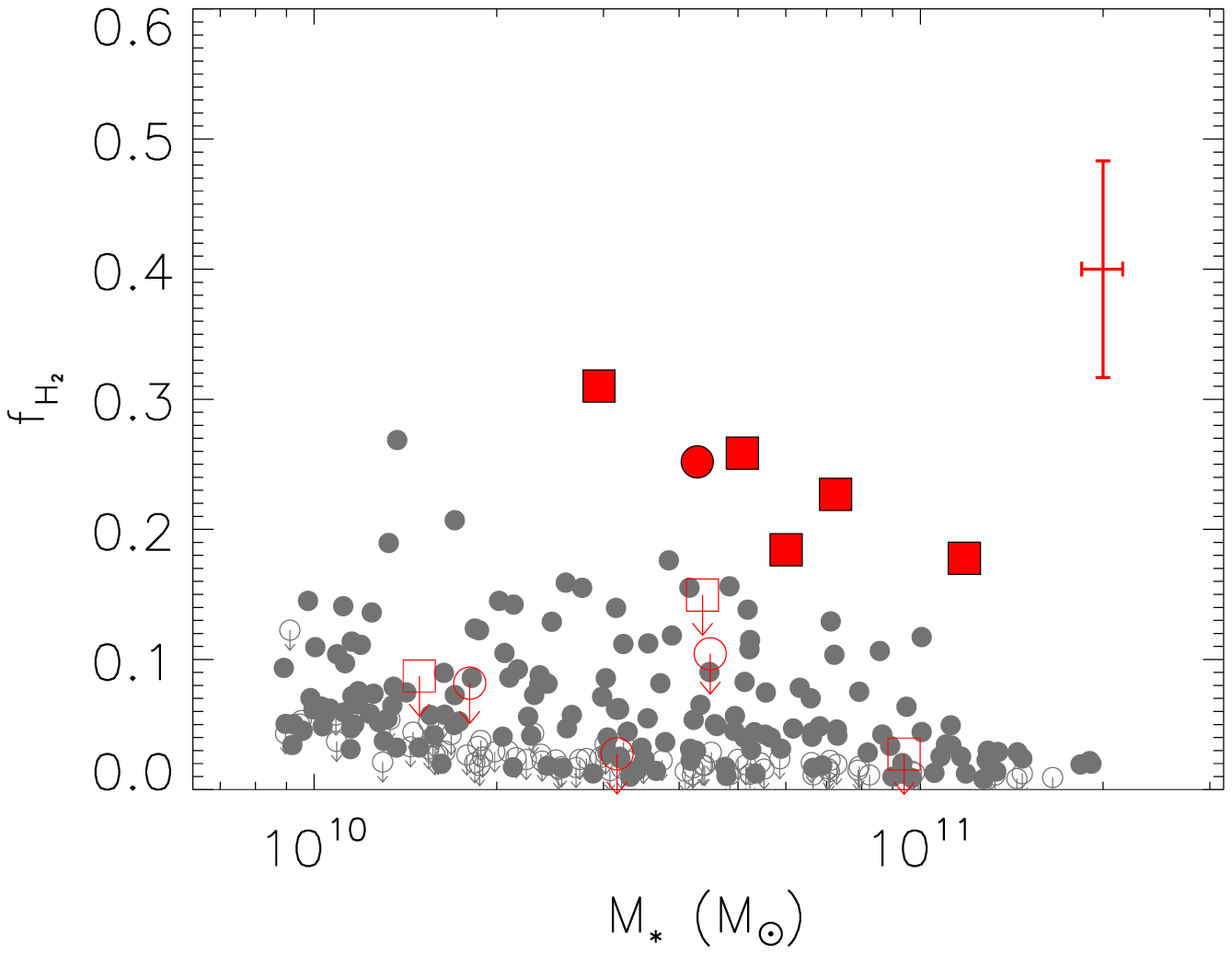}
\includegraphics[width=3.5in]{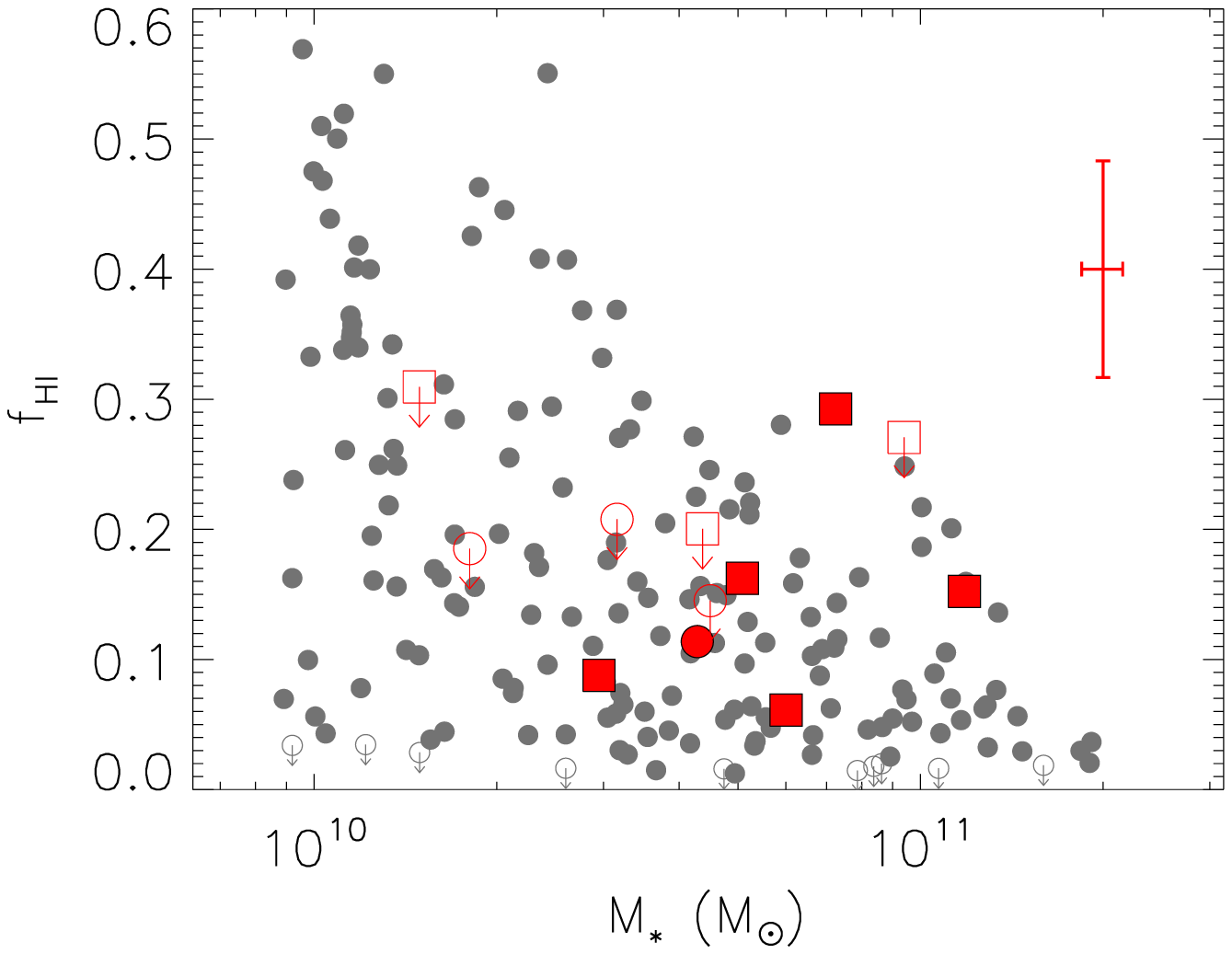}
\end{tabular}
\caption{(Left) Molecular gas fraction versus stellar mass for our target sample (red) compared to the COLD GASS sample (grey). (Right) Atomic gas fraction versus stellar mass for our target sample (red) compared with the COLD GASS sample (grey). As in previous figures, unfilled red symbols indicate non-detections.}
\label{fig:gas_fracts}
\end{figure*}

We further examine the differences in the relative quantities of atomic and molecular gas between our sample and COLD GASS by comparing their baryonic fractions of the two gas components. We calculate the fractions of molecular and atomic gas as:

\begin{equation}\label{mol_gas_fract_eq}
f_{\rm H_2} = M_{\rm H_2}/(M_{\rm H_2} + M_{\rm HI} + M_*)
\end{equation}

\begin{equation}\label{hi_gas_fract_eq}
f_{\rm HI} = M_{\rm HI}/(M_{\rm H_2} + M_{\rm HI} + M_*),
\end{equation}

and present the gas mass fractions in Figure \ref{fig:gas_fracts}. We also find that the molecular gas fractions for our targets are in excess of the majority of the COLD GASS sample, with $10-40$ per-cent molecular gas fractions for our targets that have detections in both gas components. As before, we also find that the \hi gas fractions for our sample are typical of the atomic gas fractions for the reference sample, given their stellar masses. While the panels of Figure \ref{fig:gas_fracts} show how the relative fractions of molecular and atomic gas components compare for the two samples, they do not show a direct comparison between the molecular and atomic gas masses for these samples.

\begin{figure}
\centering
\includegraphics[width=3.4in]{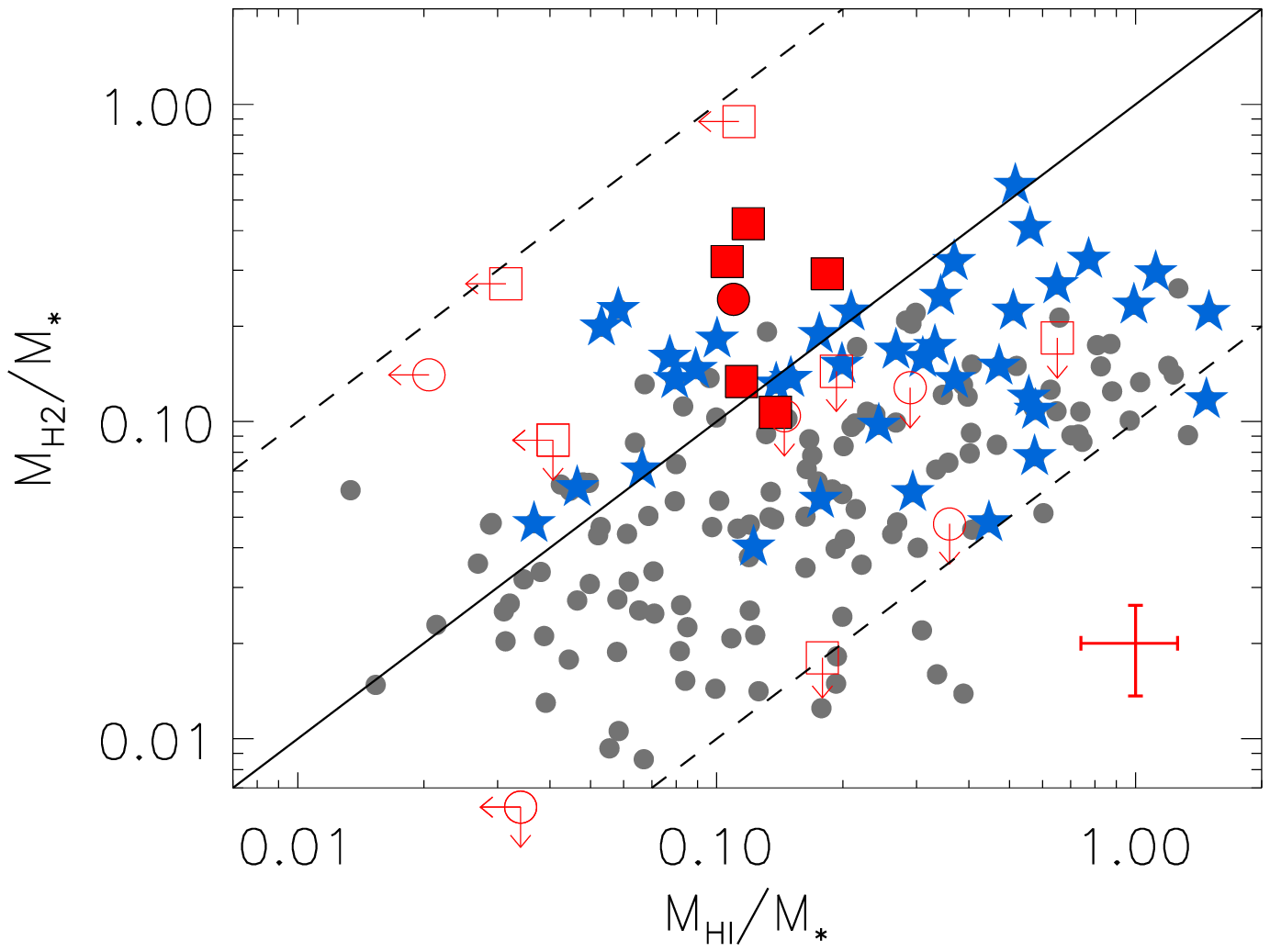} 
\caption{Molecular gas mass versus atomic gas mass, both normalized by stellar mass, for our sample (red) compared to the COLD GASS sample. The blue stars indicate galaxies in the COLD GASS sample with $L_{\rm IR}\geq10^{9.5}\lsun$, and the grey circles are galaxies in COLD GASS with $L_{\rm IR}<10^{9.5}\lsun$. The solid line indicates a 1:1 ratio of molecular-to-atomic gas mass, while the upper and lower dashed lines indicate a 10:1 and 1:10 ratio of molecular-to-atomic gas mass, respectively.}
\label{fig:mh2_vs_mhi}
\end{figure}

A direct comparison between the molecular and atomic gas masses for the two samples is shown in Figure \ref{fig:mh2_vs_mhi}. Here we find the differences in molecular-to-atomic gas between our sample and the COLD GASS reference galaxies the most apparent. The galaxies detected in both \coshort and \hi in our survey all lie on or above the 1:1 line in Figure \ref{fig:mh2_vs_mhi}, whereas the vast majority of the COLD GASS galaxies are on the \hi-dominated side. However, when we highlight the more infrared luminous galaxies in the COLD GASS sample, the $\sim15$ per cent with $L_{\rm IR}\geq10^{9.5}\lsun$, in Figure \ref{fig:mh2_vs_mhi}, we do see that the infrared-selected subset does include most of the galaxies that are more molecular gas rich in the COLD GASS sample. 

\subsection{Molecular Gas and Environment}\label{gas_and_environ}

Our sample for this pilot study is split between cluster members (12) and non-members (11), and so we can explore, in a very basic way, the differences we see between the half of our sample located in clusters to the half outside of the clusters. The general statistics for cluster members versus non-members are presented in Table \ref{tab:environ}. With such small numbers, it's difficult to draw any significant conclusion from a comparison of the cluster members versus the non-members, apart from the fact that the molecular gas detection rates seem to have no strong dependence on whether the galaxies are in a cluster or not. 

A meaningful examination of the role of environment requires a more complete sampling of the cluster environment than what we can accomplish with such a small target list. Our cluster members include just one galaxy (J101730.0+385831) having a projected radius within the virial radius of its parent cluster, although it is worth noting that this particular galaxy is detected in \coshort. Of the remaining cluster members, seven have projected radii of between 1-2 times the virial radius, and four are projected at more than 2 times the virial radius. There is no obvious correlation between projected radius and \coshort content that we can discern from this sample. A follow-up to this pilot study has already been carried out in the LMT Early Science 3 (ES3) phase that addresses the limitations of the present study, and will be presented in Cybulski et al. (in prep). More details of this follow-up study are described in the discussion at the end of Section \ref{conclusions}.

\begin{table*}
 \centering
  \caption{Basic statistics on the cluster members and non-members in our sample, including the number of detections of \coshort, and the mean signal-to-noise of those detections.} \label{tab:environ}
  \begin{tabular}{@{}lllll@{}}
  \hline
           & \# Total    &  \# Detections  &  \# Reliable Detections & $<S/N>$    \\
           &             &  (S/N$>$2.0)   &  (S/N$>$3.6)             & (detections) \\
 \hline
   &             &                 &       &      \\
 Cluster Memb.     & 12  &    8      &  5  & 4.9  \\
 Cluster Non-memb. & 11  &    7      &  4  & 6.1  \\
\hline
\end{tabular}
\end{table*}

\section{Conclusions and discussion}\label{conclusions}

We have completed a pilot \coshort study, COOL BUDHIES, targeting a sample of galaxies in and around $z\sim0.2$ clusters using the Redshift Search Receiver on the new Large Millimeter Telescope. Our sample consists of about half galaxies that are \hi selected, and half which lack \hi detections but are selected based on MIPS [24]. Out of 23 galaxies in our sample, we reliably detect the \colong emission line (with $S/N\geq3.6$) in nine galaxies, and we derive FWHM and \coshort redshifts for those galaxies. We also find marginal detections (with $2.0< S/N<3.6$) for six galaxies, which we treat differently from the non-detections because we find the emission line in the spectrum co-incident with the expected frequency based on their prior redshift information. For the remaining eight galaxies in our sample we fail to detect the \coshort line with even a marginal statistical significance in $\sim$1hr of integration. 

There is an obvious correlation between infrared luminosity and the quantity of molecular gas, consistent with previous studies \citep{youngscoville1991}. Eight out of the nine LIRGs in our sample (89 per-cent) are detected in \coshort, while only seven out of the 14 galaxies below the LIRG threshold in infrared luminosity (50 per-cent) have molecular gas detections. We also find our most molecular gas rich systems to typically be the most infrared luminous, as the subset of our targets having $M_{\rm H_2}\;\geq10^{10}\msun$ includes all but one of the LIRGs.

\begin{figure}
\centering
\includegraphics[width=3.4in]{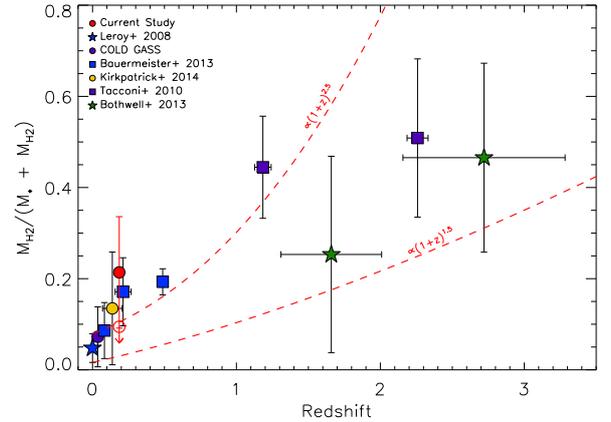} 
\caption{Molecular gas fraction (ignoring atomic Hydrogen) as a function of redshift. The median molecular gas fraction for the galaxies with \coshort detections in our current study are plotted with the red filled circle. The unfilled red circle indicates the median upper limit on the molecular gas fraction from our non-detections. Also plotted are the range of molecular gas fractions from \citet{leroy2008} (blue star), \citet{saintonge2011a} (purple circle), \citet{bauermeister2013} (blue squares), \citet{kirkpatrick2014} (yellow circle), \citet{tacconi2010} (purple squares), and \citet{bothwell2013} (green stars). The red dashed lines show the approximate upper and lower limits for the proposed redshift evolution of the molecular gas fraction from \citet{geach2011}. Note that the molecular gas fractions for our detections lie about $1\sigma$ above the upper boundary of the proposed trend.}
\label{fig:h2_fract_compare}
\end{figure}

The tendency to find our target sample more molecular gas dominated than the reference sample could be explained as being due to one, or more, reasons:

\begin{enumerate}
\item \textbf{Gas stripping in the cluster environment preferentially removing atomic gas.} Recent work in the most massive BUDHIES cluster, A963, by \citet{jaffe2015} found unmistakable evidence of stripping of \hi gas due to ram-pressure stripping (RPS) for galaxies in the parts of projected phase-space where models suggest RPS to be a significant factor. However, we expect to find that molecular gas, owing to its more central concentration in the galactic disk and its position deeper within the galactic potential, is more resilient to stripping compared to the more loosely-bound atomic gas component. Therefore, we might expect to find galaxies in cluster environments, at least those that still have cold gas, to have a higher fraction of molecular-to-atomic gas. However, many of our targets that we have found to be extremely abundant in molecular gas have been selected at redshifts slightly in the foreground or background of our clusters, and would not be subjected to the effects of RPS. Moreover, as described earlier most of our cluster member sample is not located in within the virial radii of their parent cluster, and therefore they would not be expected to have experienced significant stripping of their gas yet. Indeed, seven out of the eight A963 members from our pilot study have phase-space coordinates consistent with having recently joined the cluster, and it is therefore certain that RPS in the cluster environment is not a significant factor for them. We can also safely conclude that the influence of the cluster environment is negligible in this study, because the cluster galaxies, plotted as red squares in the right panels of Figures \ref{fig:gas_masses} and \ref{fig:gas_fracts}, are consistent with the field comparison sample of COLD GASS and show no obvious signs of being deficient in atomic gas due to their environment. It is abundantly clear, however, that we need improved statistics and a more complete sampling of phase space in the cluster environment to assess the significance of RPS on the gas content of these cluster galaxies.
\item \textbf{Redshift evolution in the molecular gas fraction} from $z\sim0$ to $z\sim0.2$. Recent work by \citet{genzel2015}, combining new molecular gas measurements with work from the literature, indicates that one should expect to find approximately a 30 per-cent increase in the molecular gas fraction for galaxies on the ``star forming main sequence'' between the redshift of the COLD GASS sample and the BUDHIES clusters. Similarly, \citet{geach2011} proposed that the molecular gas fraction evolves as $\propto(1+z)^{2\pm0.5}$, which implies an increase in the molecular gas fraction of $\sim30\pm8$ per-cent for a similar sample of galaxies between the redshift of our comparison sample and the sample in our pilot study. To examine this further, we have compiled data from various sources in the literature, consisting of molecular gas masses and stellar masses, spanning a wide redshift range to place this study in the context of our current picture of the evolution of molecular gas abundance in galaxies. We show this molecular gas abundance comparison in Figure \ref{fig:h2_fract_compare}, along with the approximate upper and lower boundaries of expected evolution based on the proportionality of \citet{geach2011}. One important caveat of Figure \ref{fig:h2_fract_compare} is that these various studies comprise a highly heterogeneous sample of galaxies, and differences intrinsic to the particular Hubble types, star-formation and gas accretion histories, environments, and ISM properties of the galaxies being sampled could all affect the abundance of molecular gas in these studies. Nevertheless, it is still relevant to place our results in context with other studies, as it can help highlight how or why our sample differs with those studies. It is fairly clear from our comparison in Figure \ref{fig:h2_fract_compare} that the relative abundance of molecular gas that we find in our study is not accounted for in terms of the expected redshift evolution alone.
\item \textbf{Differences between aperture sizes in our CO observations.} As we discussed in Section \ref{cg}, the relative aperture sizes for \coshort observations, compared to the angular sizes of our targets, differ greatly between our study and the COLD GASS study. Although \citet{saintonge2011a} presented a technique for applying an aperture correction to recover the total \coshort luminosity, including additional pointings for the most extended of their targets, our similarly-sized beam is much better suited to recover the total \coshort flux in our sample of galaxies at $z\sim0.2$ (as our aperture corrections would be on the order of 2 per-cent). However, we can conclude that the aperture differences are not the dominant cause for the contrasting molecular gas fractions in these two samples, based on the fact that our detections lie well above the upper boundary of the proposed trend in molecular gas fraction in Figure \ref{fig:h2_fract_compare}.
\item \textbf{Target selection and Eddington bias.} About half of our sample (those lacking \hi detections) are selected in the infrared, and that selection process will inherently result in a greater fraction of galaxies rich in molecular gas. However, this infrared selection alone does not account for the overall trend we have seen in our sample. For example, in Figure \ref{fig:mh2_vs_mhi} there are only six galaxies plotted that are infrared selected (those having upper limit arrows pointing to the left). One likely important effect driving the apparently high abundance of molecular gas in the targets of our study could be Eddington bias, as our limits of \coshort detection are not as low as in the COLD GASS study. Therefore, we only have reliable detections of molecular gas for the ``upper envelope'' of the most molecular gas rich subset of galaxies we have targeted (e.g., our detections are on the upper boundaries of the distribution of our comparison sample in the left panels of Figures \ref{fig:gas_masses} and \ref{fig:gas_fracts}). The significance of Eddington bias in affecting our results is probably most clear from the upper limit point shown in Figure \ref{fig:h2_fract_compare}, which shows that if we take the median upper limit on the molecular gas fraction for our non-detections, we find fractions that are in agreement with the expected redshift evolution proposed by \citet{geach2011}. We therefore conclude that the dominant cause of the unusually high molecular gas abundance seen in our pilot study is due to the Eddington bias, as sufficiently long exposure times on our non-detections would likely lead to overall molecular gas fractions that are more typical of the redshift range and infrared luminosities of our pilot study galaxies.
\end{enumerate}

One of the primary goals of the BUDHIES project is to understand the evolution of cold gas in galaxies out to intermediate redshifts, and in particular to study the effects (if any) of the cluster environment on that evolution. For our pilot COOL BUDHIES study, we have only a limited sample of targets to examine. Nevertheless, by choosing our sample strategically we have found some insight into the effects of target selection that will benefit our future study. Our sample seems to show dramatically greater evolution in the fraction of molecular gas, relative to the low-redshift comparison sample, than what is predicted by prior studies \citep[e.g.,][]{geach2011}, but much of this apparent evolution in molecular gas fraction could be attributed to Eddington bias or to the sample selection. As such, it is difficult to draw any significant conclusions about the overall evolution of cold gas in galaxies out to $z\sim0.2$ from this sample alone.  

From this study we have confirmed that infrared luminosity is an extremely effective predictor of molecular gas abundance, while \hi has little-to-no relation with $\htwo$ for our sample at $z\sim0.2$. Our follow-up study, which has already been carried out in the LMT Early Science 3 (ES3) season, has targeted an additional 43 galaxies in the cluster A963. The expanded study includes targets that populate different parts of projected phase space and a range of galaxy colours. Our combined ES2 and ES3 spectra will include 50 galaxies populating the dynamical space around the massive cluster A963, and all having a range of \hi masses and infrared luminosities. With these data we will more sensitively probe the effects of the cluster environment on molecular gas, as even with non-detections (which we anticipate for redder, less infrared-luminous galaxies) we will have sufficient statistics to stack on our \coshort spectra to examine the average molecular gas content of galaxies as a function of stellar mass, colour, infrared luminosity, and environment. Furthermore, a side-by-side comparison to stacks of \hi spectra for our galaxies will provide a much more statistically robust examination of the effects of environment on the atomic and molecular component of the ISM. The observations have been completed for the follow-up ES3 study, and reductions of the spectra are finished. The results will be presented in a forthcoming paper.

\section*{Acknowledgements}

This research has made use of the NASA/IPAC Extragalactic Database (NED) which is operated by the Jet Propulsion Laboratory, California Institute of Technology, under contract with the National Aeronautics and Space Administration. This research has also made use of NASA's Astrophysics Data System (ADS), and the NASA/IPAC Infrared Science Archive, which is operated by the Jet Propulsion Laboratory, California Institute of Technology, under contract with the National Aeronautics and Space Administration. This research made use of Montage, funded by the National Aeronautics and Space Administration's Earth Science Technology Office, Computation Technologies Project, under Cooperative Agreement Number NCC5-626 between NASA and the California Institute of Technology. Montage is maintained by the NASA/IPAC Infrared Science Archive. 

This publication makes use of data products from the Wide-field Infrared Survey Explorer, which is a joint project of the University of California, Los Angeles, and the Jet Propulsion Laboratory/California Institute of Technology, funded by the National Aeronautics and Space Administration. 

The United Kingdom Infrared Telescope (UKIRT) is supported by NASA and operated under an agreement among the University of Hawaii, the University of Arizona, and Lockheed Martin Advanced Technology Center; operations are enabled through the cooperation of the Joint Astronomy Centre of the Science and Technology Facilities Council of the U.K. When some of the data reported here were acquired, reported here were acquired, UKIRT was operated by the Joint Astronomy Centre on behalf of the Science and Technology Facilities Council of the U.K. 

Some of the data presented in this paper were obtained from the Mikulski Archive for Space Telescopes (MAST). STScI is operated by the Association of Universities for Research in Astronomy, Inc., under NASA contract NAS5-26555. Support for MAST for non-HST data is provided by the NASA Office of Space Science via grant NNX09AF08G and by other grants and contracts. 

Funding for the SDSS and SDSS-II has been provided by the Alfred P. Sloan Foundation, the Participating Institutions, the National Science Foundation, the U.S. Department of Energy, the National Aeronautics and Space Administration, the Japanese Monbukagakusho, the Max Planck Society, and the Higher Education Funding Council for England. The SDSS Web Site is http://www.sdss.org/.

R. Cybulski was supported by NASA Contract 1391817 and NASA ADAP grant 13-ADAP13-0155. M.~S. Yun acknowledges support from the NASA ADAP grant NNX10AD64G.




\bibliographystyle{mnras}
\bibliography{thebib} 



\appendix \label{appendix}

\section{\coshort Spectra} \label{allspectra}

\begin{figure*}
	\includegraphics[width=6.4in]{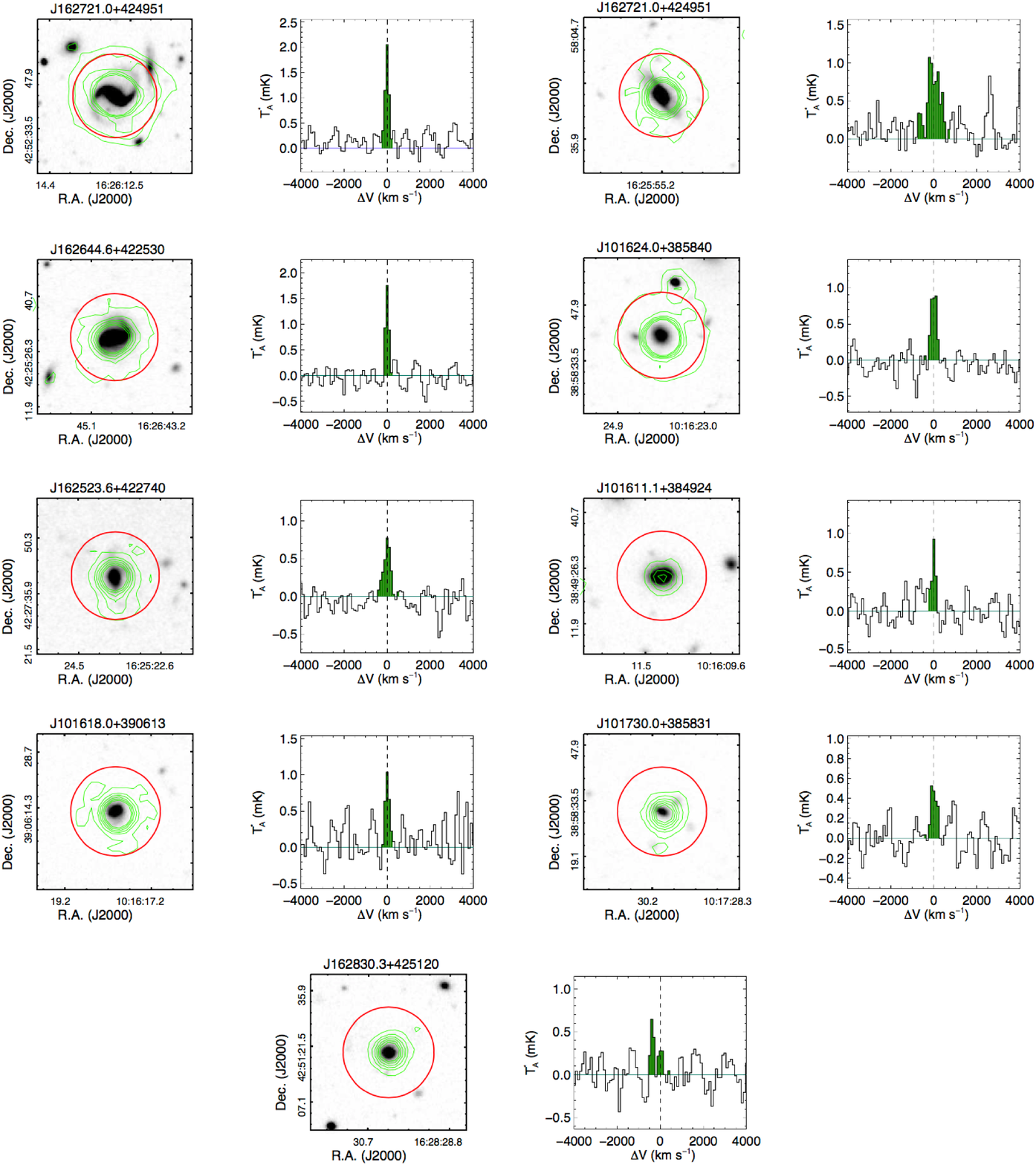}
	
	\caption{Each pair of panels corresponds to one of our target galaxies with a reliable detection of the \coshort line. The left panels show the INT R-band maps centered on the galaxy. The red circle shows the RSR beam. In green we plot 5-30$\sigma$ (in intervals of 5$\sigma$) contours from MIPS [24]. The right panels give the spectrum of the galaxy showing the interval of $\pm$4000km s$^{-1}$ centered on the velocity of the \coshort line.}
	\label{fig:spectra_strong_detections}
\end{figure*}

\begin{figure*}
	\includegraphics[width=6.4in]{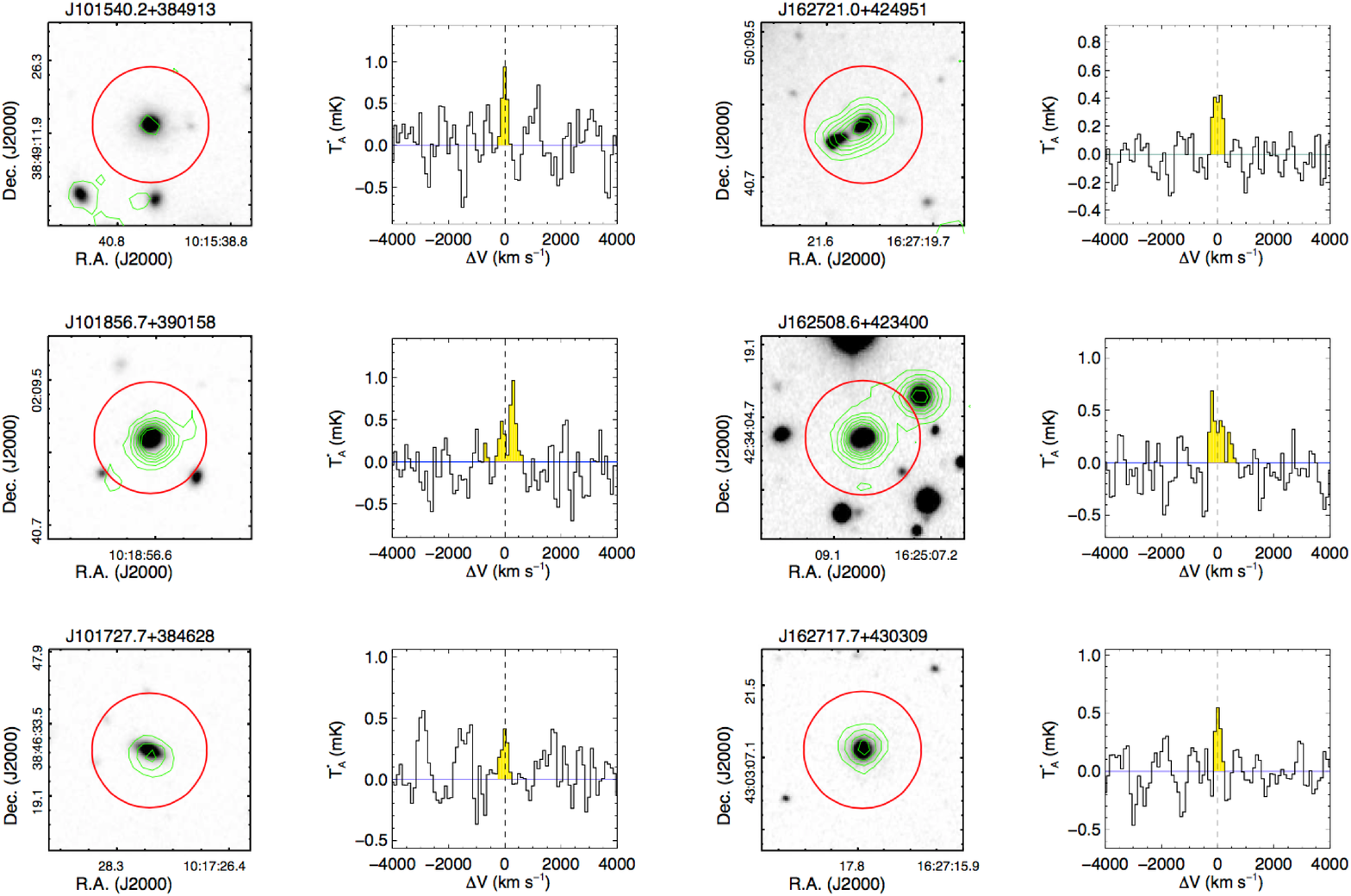}

	\caption{Each pair of panels corresponds to one of our target galaxies with a marginal detection of the \coshort line. The left panels show the INT R-band maps centered on the galaxy. The red circle shows the RSR beam. In green we plot 5-30$\sigma$ (in intervals of 5$\sigma$) contours from MIPS [24]. The right panels give the spectrum of the galaxy showing the interval of $\pm$4000km s$^{-1}$ centered on the velocity of the \coshort line.}
	\label{fig:spectra_marginal_detections}
\end{figure*}

\begin{figure*}
	\includegraphics[width=6.4in]{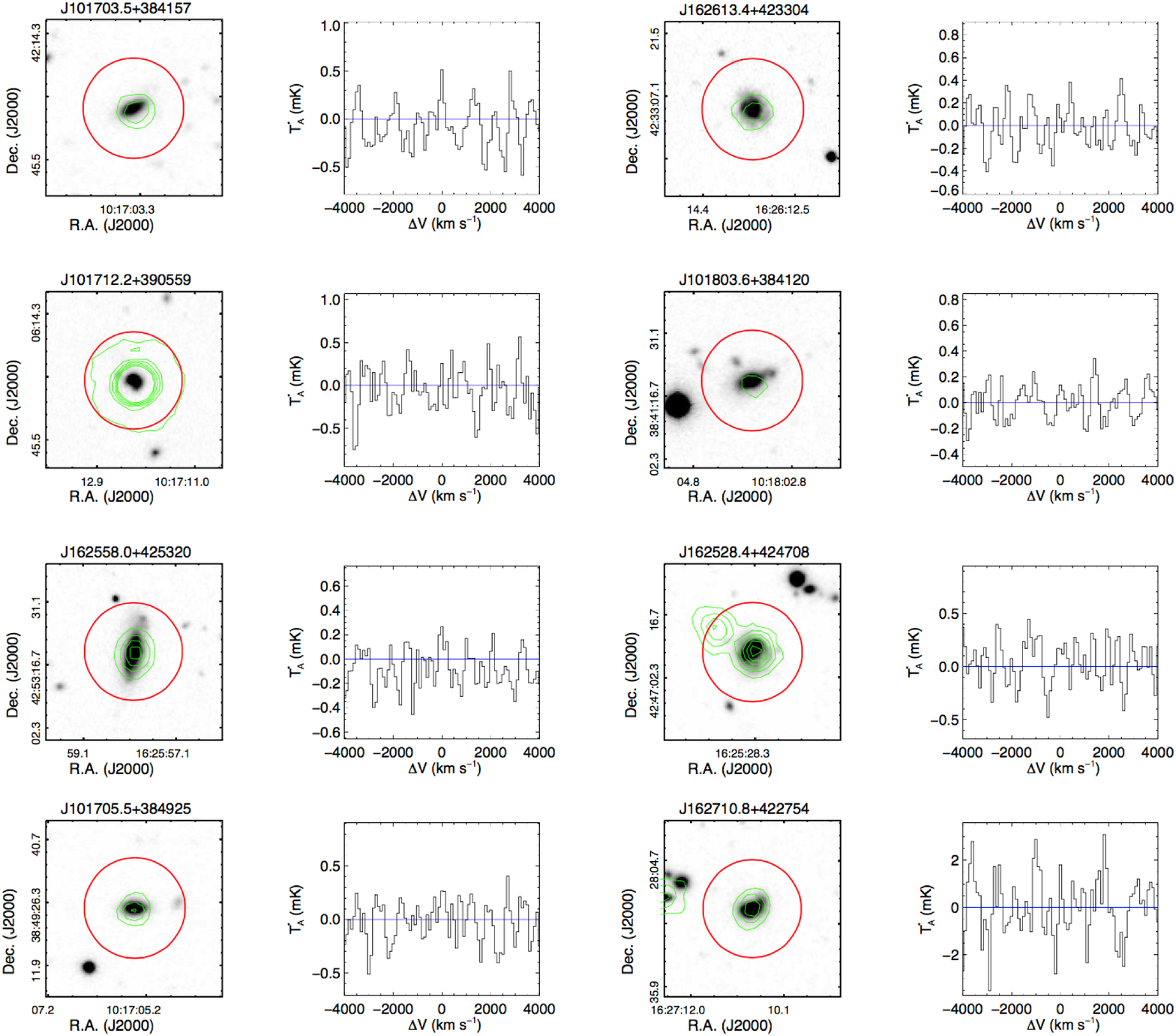}
	
	\caption{Each pair of panels corresponds to one of our target galaxies with a non-detection of the \coshort line. The left panels show the INT R-band maps centered on the galaxy. The red circle shows the RSR beam. In green we plot 5-30$\sigma$ (in intervals of 5$\sigma$) contours from MIPS [24]. The right panels give the spectrum of the galaxy showing the interval of $\pm$4000km s$^{-1}$ centered on the velocity of the \coshort line.}
	\label{fig:spectra_non_detections}
\end{figure*}

Here we present the spectra for all of our targets, sorted by decreasing signal-to-noise detections of the \coshort line and grouped into reliable detections (Figure \ref{fig:spectra_strong_detections}), marginal detections (Figure \ref{fig:spectra_marginal_detections}), and non-detections (Figure \ref{fig:spectra_non_detections}).

\begin{figure*}
	\includegraphics[width=6.0in]{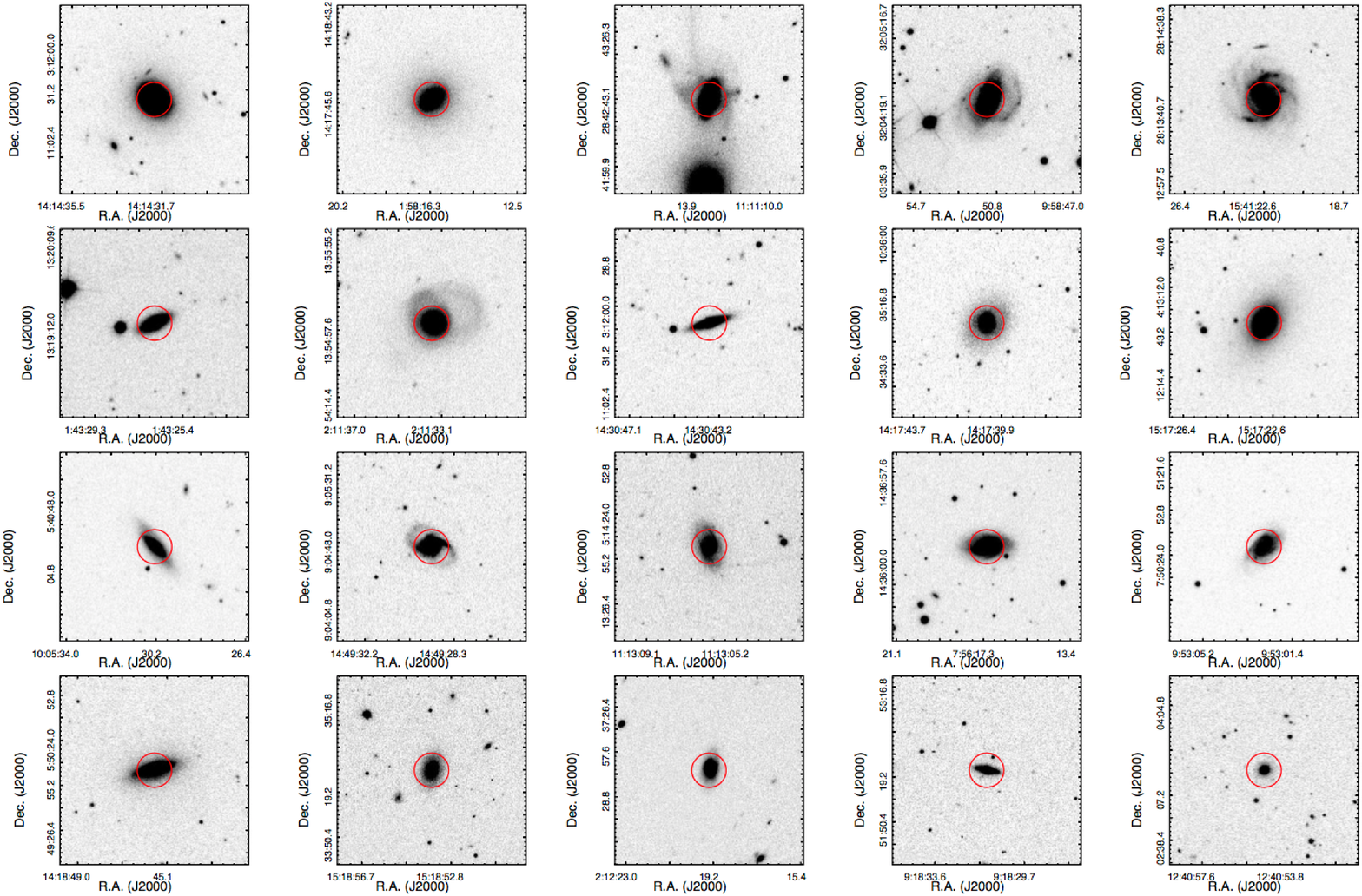}
	\caption{Random selection of galaxies from the COLD GASS survey. The images show the SDSS $r^\prime$-band, and the red circles are the 22$\arcsec$ beam FWHM for \coshort observations in the COLD GASS study.}
	\label{fig:coldgass_beam_comparison}
\end{figure*}


\bsp	
\label{lastpage}
\end{document}